\documentclass[trackchanges]{aastex701}

\usepackage{amsmath}
\usepackage[T1]{fontenc}
\usepackage{appendix}

\begin{document}

\title{Numerical Simulations of the Circularized Accretion Flow in Population III Star Tidal Disruption Events. II. Radiative Properties}

\author[orcid=0009-0008-9058-7023,gname=Yu-Heng,sname='Sheng']{Yu-Heng Sheng}
\affiliation{Astrophysics Division, Shanghai Astronomical Observatory, Chinese Academy of Sciences, 80 Nandan Road, Shanghai 200030, People’s Republic of China}
\affiliation{School of Astronomy and Space Sciences, University of Chinese Academy of Sciences, 19A Yuquan Road, Beijing 100049, China}
\email[]{shengyuheng@shao.ac.cn} 

\author[orcid=0000-0002-0427-520X,gname=De-Fu, sname='Bu']{De-Fu Bu} 
\affiliation{Shanghai Key Lab for Astrophysics, Shanghai Normal University, 100 Guilin Road, Shanghai 200234, China}
\email[show]{De-Fu Bu (dfbu@shnu.edu.cn)}

\author[orcid=0000-0002-1908-0536, gname=Liang, sname='Chen']{Liang Chen}
\affiliation{Astrophysics Division, Shanghai Astronomical Observatory, Chinese Academy of Sciences, 80 Nandan Road, Shanghai 200030, People’s Republic of China}
\email[]{chenliang@shao.ac.cn} 

\author[orcid=0000-0002-3073-5871, gname=Shi-Yin, sname='Shen']{Shi-Yin Shen}
\affiliation{Astrophysics Division, Shanghai Astronomical Observatory, Chinese Academy of Sciences, 80 Nandan Road, Shanghai 200030, People’s Republic of China}
\email[]{ssy@shao.ac.cn}

\author[0009-0008-0518-6795]{Bo-Yan Chen}
\affiliation{State Key Laboratory of Particle Astrophysics, Institute of High Energy Physics, Chinese Academy of Sciences, Beijing 100049, People's Republic of China}
\affiliation{University of Chinese Academy of Sciences, Chinese Academy of Sciences, Beijing 100049, People's Republic of China}
\email[]{bychen@ihep.ac.cn}

\author[gname=Xiao-Hong, sname='Yang']{Xiao-Hong Yang} 
\affiliation{Department of Physics, Chongqing University, Chongqing 400044, China }
\email[]{yangxh@cqu.edu.cn}

\begin{abstract}

Tidal Disruption Events (TDEs) release enormous amounts of energy, offering a promising avenue for detecting Population III (Pop III) stars. However, the radiative properties of TDEs of Pop III stars have so far been studied only analytically, relying on many assumptions. Based on our radiative hydrodynamic simulations that follow the 
evolution of the accretion system for Pop III star TDEs where a $300\ M_{\odot}$ ($M_{\odot}$ is the solar mass) star is disrupted by a $10^{6}\ M_{\odot}$ black hole (BH), we compute the emission properties of the event in rest frame and find that the spectrum peaks in the optical/UV waveband. After accounting for redshift ($z \sim 10$) and extinction effects, we find the observed spectral peak shifts to the infrared, with fluxes exceeding $10^{2}~\mathrm{nJy}$—making such events detectable with both the James Webb Space Telescope (JWST) and the Nancy Grace Roman Space Telescope (Roman). The dependence of the observed spectrum on viewing angle is suppressed due to dust extinction. Using our simulation results, we also calculate the radio emission generated by the interaction between the wind and the circumnuclear medium (CNM) and find that a Pop III star TDE can produce an unusually long-lasting, continuously increasing radio flare with a duration greater than $10^4$ days and thus has the potential to be detected in radio wavebands. These results may be helpful to the detection of Pop III stars.

\end{abstract}



\section{INTRODUCTION} 

A tidal disruption event (TDE) occurs when a star passes close to a central supermassive black hole (SMBH) and its pericenter distance ($R_{\rm p}$) is less than or equal to the tidal disruption radius ($R_{\rm T}$) (\cite{1975Natur.254..295H}; \cite{Rees1988}; \cite{Evans1989}). The star is then torn apart by the SMBH's tidal forces. Approximately half of the stellar debris becomes unbound and is ejected, while the remaining portion falls back and eventually accretes onto the SMBH. The temporal evolution of the fallback rate follows $\dot{M}_{\rm fb} \propto t^{-5/3}$ (\cite{Evans1989}; \cite{Phinney1989}; \cite{Lodatoetal2009}; \cite{GuillochonRamirezRuiz2013}). The peak injection rate could exceed $10^{26}\rm\ g\ s^{-1}$ for a typical TDE in which a solar-type star is disrupted by a $10^{6-7} M_{\odot}$ SMBH. Such a high rate of mass injection can establish a super-Eddington accretion system, which releases a considerable amount of radiative energy. For a typical TDE, the resulting accretion flow is expected to have a size of $\sim 10^{13}$ cm and a temperature of $\sim 10^5$ K (see \cite{2020SSRv..216..114R} for a review). Consequently, the emission spectrum of such a system is expected to peak in the soft X-ray band (\cite{Cannizzoetal.1990}; \cite{Ulmer1999}; see \cite{2020SSRv..216..114R} for reviews).

TDEs were indeed identified in the soft X-ray band by observations in the 1990s (see \cite{Komossa2015} for a review). However, in the past decades, a distinct class of TDEs has been discovered in the Optical/UV bands. The inferred radiation size of these events ($\sim10^{14}-10^{16}$ cm) is significantly larger than the theoretical predictions (\cite{Hungetal2017}; \cite{vanVelzenet42al.2020}; \cite{Gezari2021}) and the radiation temperature of them is much lower (\(\sim\!10^{4}\) K; e.g., \cite{Gezarietal.2012}; \cite{Arcavietal.2014)}). Recent studies have proposed two main scenarios to explain this phenomenon. The first scenario suggests that shocks are produced by the collision of debris streams during fallback (\cite{Guillochon&Ramirez-Ruiz2015}; \cite{Shiokawaetal.2015}). These shocks dissipate energy and power the observed Optical/UV emission (\cite{Piranetal.2015}; \cite{JiangGuillochonLoeb2016}; \cite{Steinberg&Stone2024}; \cite{2025ApJ...979..235G}). The second scenario posits that an accretion disk, undergoing a super-Eddington accretion rate, develops an optically thick envelope. Some studies also suggest that the super-Eddington accretion disk launches a powerful wind, which is found in recent observations (e.g., \cite{2017ApJ...846..150Y}; \cite{2018MNRAS.474.3593K}; \cite{2020MNRAS.494.4914P}; \cite{2025ApJ...989L...9L}; \cite{2019ApJ...879..119H}; \cite{2024ApJ...972..106X}) . The high-energy photons generated near the SMBH are then reprocessed into Optical/UV photons as they travel through this expanding, optically thick envelope (\cite{Loeb&Ulmer1997}; \cite{Coughlin&Begelman2014}; \cite{Rothetal.2016}; \cite{Liuetal.2017}; \cite{Liuetal.2021}; \cite{Metzger&Stone2017}; \cite{Metzger2022}; \cite{Weversetal.2022}) or within the wind itself (\cite{2009MNRAS.400.2070S}; \cite{Lodato&Rossi2011}; \cite{Metzger&Stone2016}; \cite{Piro&Lu2020}; \cite{Uno&Maeda2020}; \cite{BU22}; \cite{Parkinsonetal.2022}; \cite{MageshwaranShawBhattacharyya2023}; \cite{Curd&Narayan2019}; \cite{Dai2018}; \cite{Thomsen2022}).

In addition, nonthermal emission has also been detected in some TDEs. Two types of radio flares have been detected: prompt radio flare has been found within weeks after the main flare of the optical peak (e.g., \cite{2016ApJ...819L..25A}; \cite{2021ApJ...919..127C};
\cite{2022MNRAS.511.5328G}; \cite{2023MNRAS.522.5084G}; \cite{2023MNRAS.518..847G}; \cite{2024MNRAS.528.7123G}; \cite{2024ApJ...973..104D}) and delayed radio emission has been observed months to years after optical discovery (\cite{2020SSRv..216...81A, 2021NatAs...5..491H, 2021ApJ...920L...5H, 2022ApJ...938...28C, 2024ATel16650....1C, 2022ApJ...925..143P, 2022ApJ...933..176S, 2024ApJ...962L..18Z, 2023MNRAS.520.2417W, 2025SciA...11y9068W}). This emission is commonly attributed to synchrotron radiation from cosmic-ray electrons accelerated in forward shocks produced by the interaction of jets or winds with the circumnuclear medium (CNM) \cite{2013ApJ...772...78B, 2021MNRAS.507.4196M, 2024ApJ...971...49M, 2024ApJ...971..185C, 2025arXiv251014715M, 2025ApJ...988L..24H, 2025arXiv251223252S, 2025arXiv251221669M, 2023ApJ...954....5H, 2025arXiv251114008W}, or in bow shocks generated by their interaction with dense clouds in the vicinity \cite{Mou2022, 2023MNRAS.521.4180B, 2025ApJ...979..109Z, 2025arXiv251025033M, 2025ApJ...993L...2Y}. Additionally, other studies suggest that cosmic-ray electrons may originate from shocks produced by the interaction of unbound debris with the CNM (\cite{2016ApJ...827..127K, 2019MNRAS.487.4083Y}) or with a torus (\cite{2024ApJ...977...63L}). Although the precise origin of these cosmic-ray electrons remains unclear, the formation of a forward shock from the outflow-CNM interaction is considered inevitable if a fast outflow occurs.

TDEs release a large amount of radiative energy, making them potential probes for distant objects such as the first generation of stars. These stars, known as Population III (Pop III) stars, are believed to have formed from pristine gas a few hundred million years after the Big Bang. They typically formed in the early universe at redshifts of $z \sim 10$--$15$ (\cite{2002Sci...295...93A, 2002ApJ...564...23B, 2006ApJ...652....6Y}), with masses ranging from $30M_\odot$ to $300M_\odot$. Metals produced in the cores of Pop III stars can be released into the intergalactic medium (IGM) through supernova explosions \cite{2024ApJ...964...91C}, a process that plays an important role in shaping the subsequent formation of higher-metallicity Pop II and Pop I stars. However, directly observing Pop III stars remains a challenge. So far, only a few Pop III-like star candidates have been identified (\cite{2020MNRAS.494L..81V, 2022Natur.603..815W}). Recently, \cite{ChowdhuryChang&Daietal.2024} demonstrated that for a $10^6\ M_{\odot}$ SMBH, TDEs of Pop III stars can have a peak fallback rate of $\sim 10^{4-6}\ \dot{M}_{\rm Edd}$, where $\dot{M}_{\rm Edd} = L_{\rm Edd} / (\eta c^2)$ is the Eddington accretion rate, with $L_{\rm Edd}$ being the Eddington luminosity and $\eta$ the radiative efficiency. Such an extremely high accretion rate can produce a significant photosphere detectable in the infrared waveband even at $z \sim 10$, suggesting that TDEs are a promising method for detecting Pop III stars.

While the detailed geometry of the wind and accretion flow remains uncertain, the analytical study by \cite{ChowdhuryChang&Daietal.2024} adopted the model proposed by \cite{2009MNRAS.400.2070S}, which treats the photosphere of a Pop III star TDE as a spherically symmetric expanding layer. However, the geometry of the accretion flow can significantly influence the reprocessing of photons. Recent simulations have shown that a funnel (In this work, the term ‘funnel’ refers to the optically thin region between the photosphere and the rotational axis.) can form in super-Eddington accretion systems, allowing some X-ray photons to escape directly (\cite{Dai2018, 2019ApJ...880...67J, Thomsen2022}). Therefore, the structure of the accretion flow directly affects the observed spectrum, highlighting the necessity of studying Pop III star TDE systems within a numerical simulation framework.

There are many other simulations about typical TDEs (e.g.,\cite{2005ApJ...628..368O}; \cite{Dai2018}; \cite{Curd&Narayan2019}; \cite{Curd&Narayan2023}; \cite{Thomsen2022}; \cite{BU23}), while there are no other works, to our knowledge, simulated the structure  and evolution of the accretion system of Pop III star TDE. We conducted radiative hydrodynamic simulations in our previous work to study the evolution of Pop III star TDE accretion systems \cite{2025arXiv251221500S}. Specifically, we simulated the accretion flow resulting from the disruption of a $300M_{\odot}$ star by a $10^6M_{\odot}$ SMBH. We assumed an efficient circularization process and considered two different metallicity models: $Z = 10^{-9}Z_{\odot}$ (Model M300-9) and $Z = 10^{-5}Z_{\odot}$ (Model M300-5). The peak accretion rate reaches $\sim 5\times10^4\dot{M}_{\rm Edd}$ for Model M300-9 and $\sim 1\times10^4\dot{M}_{\rm Edd}$ for Model M300-5. In that work, we primarily focused on the properties of the wind and the accretion structure under such high accretion rates. We found that the structure of the photosphere evolves from a vertically elongated morphology to a  horizontally elongated one over time. The funnel region diminishes at early times and later re-forms. This evolution affects the resulting light curves, and the non-spherically symmetric photosphere introduces a viewing-angle dependence in the observed spectrum.

In this study, we calculated the emission properties of Pop III star TDEs based on our simulation results. We computed the blackbody emission from the photosphere to investigate the viewing-angle dependence of the spectral energy distribution and derived the observed spectrum and light curves for a source at redshift $z \sim 10$. To model observational conditions more realistically, we applied a specific extinction model that includes absorption by the IGM and extinction from a diffuse gaseous environment. Furthermore, to enable multi-wavelength observational predictions, we computed the radio flare generated by the interaction between the  wind and CNM, utilizing the wind properties derived from our simulations. The corresponding radio light curves were produced with cosmological redshift effects fully taken into account.

The structure of this paper is as follows. In Section 2, we describe the methodology used in our calculations. The results are presented in Section 3. Finally, we summarize and discuss our findings in Section 4.

\section{METHOD}
\subsection{Blackbody emission}
\begin{figure}
    \centering
    \includegraphics[width=0.7\linewidth]{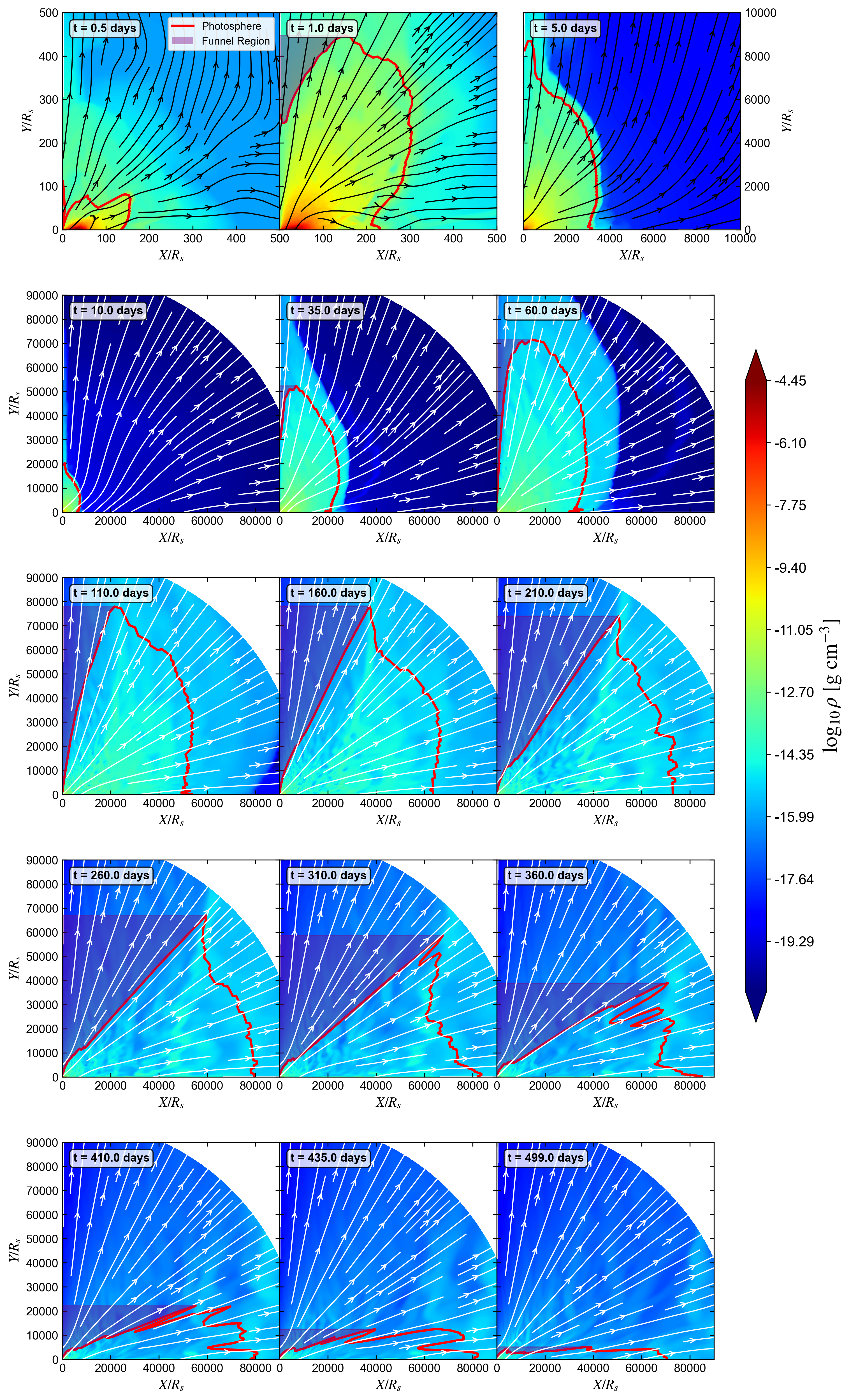}
    \caption{Snapshots of gas density (colour scale) and fluid velocity streamlines (black arrows are used for the panels in the first row, and white arrows for the other panels.) for model M300-9 at 15 evolutionary stages: from $t = 0.5\,\mathrm{days}$ (top-left panel), to  $t = 499.0\,\mathrm{days}$ (bottom-right panel). The electron scattering photosphere is indicated by the red contour and the funnel region is denoted as purple shade in corresponding panels. The first two panels in the top row concentrate the zoom-in region $0\le r\le500 R_{\rm s}$; the third panel in the same row covers $0\le r\le10000 R_{\rm s}$, and the remaining panels extend to $0\le r\le90000 R_{\rm s}$.}
    \label{fig1}
\end{figure}

We describe the method used to calculate the blackbody emission of the system. Our previous work \cite{2025arXiv251221500S} obtained the evolution of the accretion flow in Pop III star TDEs. Figure~\ref{fig1} shows snapshots of the gas density and streamlines for the system. The electron scattering optical depth is defined as
\[
\tau(\theta, r) = \int_{r}^{10^{5} R_{\mathrm{s}}} \rho \kappa_{\mathrm{es}}  \mathrm{d} r^{\prime},
\]
where \(\kappa_{\mathrm{es}}\) is the electron scattering opacity. Its value lies between \(\kappa_{\mathrm{es}} = 0.34\ \mathrm{cm^{2}\,g^{-1}}\)  and \(\kappa_{\mathrm{es}} = 0.35\ \mathrm{cm^{2}\,g^{-1}}\) depending on the metallicity. We have calculated results using either value and found that it yields negligible differences in our results; therefore, for simplicity, we adopt \(\kappa_{\mathrm{es}} = 0.34\ \mathrm{cm^{2}\,g^{-1}}\) in this work. \(\rho\) is the gas density, and \(10^{5} R_{\mathrm{s}}\) is the outer boundary of the simulation region. Using the gas density distribution, we calculated the location of the photosphere ($\tau=1$) at each snapshot.

The simulation performed in \citet{2025arXiv251221500S} is two-dimensional and assumes axisymmetry, covering only the region above the mid-plane. To generate a three-dimensional photosphere, we rotate the 2D result about the polar axis and then reflect it across the mid-plane to obtain the full $0 \le \theta \le \pi$ region. The resulting photosphere is divided into $2 \times 128 \times 360$ surface elements. To investigate the viewing-angle dependence of the spectrum, we select five lines of sight ($\theta_{\rm obs} = 1^\circ, 30^\circ, 45^\circ, 60^\circ, 90^\circ$) for each simulation snapshot. For a given viewing direction, we calculate the mutual obstruction among the surface elements to identify those visible from that direction. Specifically, for a given observing direction, we consider each surface element on the photosphere. We trace a ray from that element along the line of sight. If this ray intersects any other part of the photosphere, it implies that the radiation originating from the initial surface element would have to pass through the optically thick interior of the photosphere. Given that the photosphere in our simulation is extremely geometrically thick, we assume such radiation becomes trapped and cannot reach the distant observer directly. Therefore, we define such surface elements as being obscured (and thus invisible) along that particular line of sight at that time.

The specific flux along the line of sight is then computed as
\begin{equation}
    F_{\nu}(\nu, z,\theta) = \frac{1+z}{4\pi D_{\rm L}^2} \int_{\mathrm{visible}} B_{\nu}[(1+z)\nu, T] \, \cos\vartheta \,  dA,
    \label{eq1}
\end{equation}
where $B_{\nu}$ is the blackbody intensity, $z$ is the source redshift, and $T$ and $dA$ are the temperature and area of each visible surface element, respectively. The angle $\vartheta$ is measured between the normal vector of a surface element and the line of sight. The luminosity distance $D_{\rm L}$ is given by
\begin{equation}
    D_{\rm L} = (1+z)\,c \int_0^z \frac{dz'}{H(z')},
    \label{eq2}
\end{equation}
where $c$ is the speed of light and $H(z)$ is the Hubble parameter. In this work, we adopt a flat $\Lambda$CDM cosmological model, for which the Hubble parameter is
\begin{equation}
    H(z) = H_0 \sqrt{(1+z)^3 \Omega_{\mathrm{m}0} + \Omega_{\Lambda0}},
    \label{eq3}
\end{equation}
with the parameter values $H_0 = 70~\mathrm{km~s^{-1}~Mpc^{-1}}$, $\Omega_{\mathrm{m},0} = 0.27$, and $\Omega_{\Lambda,0} = 0.73$.

The \textit{apparent} luminosity along a specific line of sight is then obtained as
\begin{equation}
    L_{\nu,\,\mathrm{apparent}}(\nu,\theta) = \frac{4\pi D_{\rm L}^2}{1+z} \, F_{\nu}\!\left(\frac{\nu}{1+z},\,z,\theta\right).
    \label{eq4}
\end{equation}
This \textit{apparent} luminosity, derived from the specific flux in a given viewing direction under the assumption of a spherically symmetric radiation field, does not represent the intrinsic bolometric luminosity of the entire source. Rather, it corresponds to the luminosity contributed only by the surface elements visible from that particular direction, thereby capturing the viewing-angle dependence of the emission. In this work, we adopt $z = 10$ to compute the specific flux and subsequently derive the apparent luminosity for the five chosen viewing angles. This approach allows the apparent luminosity to be readily used for estimating the received flux at other redshifts.

\subsection{Extinction}
\label{sectionextinction}
We account for two sources of extinction along the photon path from $z=10$ to $z=0$: the diffuse IGM and dust.

Observations of high-redshift quasars have revealed significant soft X-ray absorption. \cite{2011ApJ...734...26B} found that the optical depth $\tau$ at 0.5 keV shows little evolution with redshift, particularly at high $z$, and proposed that a cold (neutral) diffuse IGM model can explain the observed high-$z$ X-ray absorption. This model was later revisited by \cite{2013MNRAS.431.3159S}, who replaced the cold IGM with a Warm–Hot Intergalactic Medium (WHIM). Since the diffuse IGM is likely to be largely ionized  (e.g.
\cite{1965ApJ...142.1633G}; \cite{2001ApJ...552..473D}), in this work, we adopt the absorption model from \cite{2013MNRAS.431.3159S}, which is described by the following expression:
\begin{equation}
    \tau_{\mathrm{IGM}}(\nu, z, Z) = \frac{n_0 c}{H_0} \int_0^z \left[\sigma_{\mathrm{HHe}}(\nu(1+z')) + \sigma_{\mathrm{metals}}(\nu(1+z'), Z)\right] \frac{(1+z')^2}{\sqrt{(1+z)^3 \Omega_{\mathrm{m}0} + \Omega_{\Lambda0}}}  dz',
    \label{eq5}
\end{equation}
where $n_0 = 1.7 \times 10^{-7}~\mathrm{cm}^{-3}$ is the IGM number density at $z=0$ as derived by \cite{2011ApJ...734...26B}, and $Z$ is the metallicity. Here, $\sigma_{\mathrm{HHe}}$ denotes the photoionization cross section per hydrogen atom due to hydrogen and helium, while $\sigma_{\mathrm{metals}}$ represents the corresponding cross section contributed by metals.

\begin{figure}
    \centering
    \includegraphics[width=1.0\linewidth]{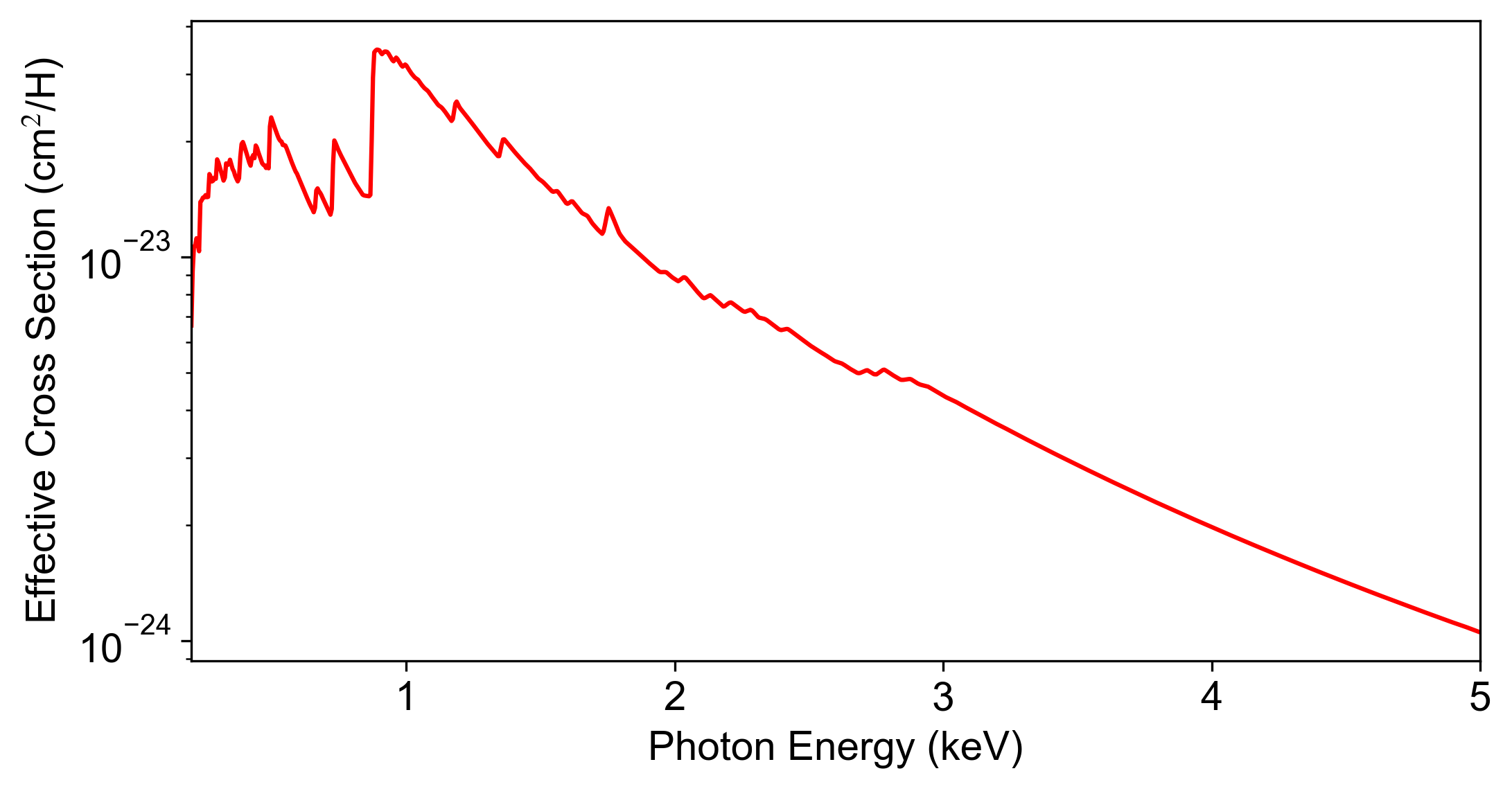}
    \caption{Photoionization cross section per atom at different photon energy for the WHIM at a temperature of $10^6$ K and a metallicity of $Z = 0.5 Z_{\odot}$, calculated using the parameters and formalism of the \textsc{xspec} model ABSORI.}
    \label{fig2}
\end{figure}

\cite{2013MNRAS.431.3159S} adopted two constant mean metallicities, $Z = 0.2Z_{\odot}$ and $Z = Z_{\odot}$, for the IGM, where $Z_{\odot}$ is the solar metallicity. They found that an IGM temperature in the range of $10^5$--$10^{6.5}$ K can account for most of the observed excess X-ray column density measured in moderate to high-redshift GRBs with metallicities $0.2Z_{\odot} \le Z \le 1Z_{\odot}$, even for GRBs at redshifts as high as $z \sim 10$. We adopted a compromise value of $Z = 0.5Z_{\odot}$ and set the IGM temperature to $10^6$ K. This prescription (\(Z=0.5Z_{\odot}\)) might overestimate the IGM metallicity in our low-metallicity Pop III context. We have also calculated results using an evolving metallicity model, \(Z(z) = 0.2Z_{\odot} * (1+z)^{-3}\), and find that they do not differ significantly from those using a constant \(Z = 0.5Z_{\odot}\). Figure~\ref{fig2} shows the cross section generated using the \textsc{xspec} model ABSORI based on \cite{1992ApJ...395..275D} with these parameters.

Besides the IGM, dust is another significant source of extinction that modifies the spectrum in the optical/UV wavebands. Recent observations of galaxies at $z\geq10$ suggest that dust would be produced by supernovae at an early time. These dusts may be largely destroyed or expelled at z>10. A marked increase in dust abundance around $z \sim 10$ could result from the fallback of previously evacuated material (see, e.g., \cite{2025Galax..13...64S} for a review). Thus for our source sited at z~10 we need to account for the dust extinction. The redder spectra observed towards higher redshifts were attributed to the effect of cosmic dust \cite{1981ApJ...250....1W}, with subsequent works further investigating cosmic dust extinction (e.g., \cite{2009ApJ...696.1727M}; \cite{2010MNRAS.405.1025M}). Specifically, \cite{2015ApJ...802L..16X} found that a constant comoving effective dust density, $n\sigma_\nu = 2\times10^{-5} h \ \mathrm{Mpc}^{-1}$, provides a good description, where $\sigma_\nu$ is the cross section of the absorber at $5500 \ \mathring{\mathrm{A}}$. The corresponding extinction model is described by the following equations:
\begin{equation}
    \tau_{\mathrm{dust}}(\nu, z) = \frac{c}{H_0} \int_0^z n\sigma_{\mathrm{dust}}(\nu(1+z')) \frac{(1+z')^2}{\sqrt{(1+z')^3 \Omega_{\mathrm{m},0} + \Omega_{\Lambda,0}}}  dz',
    \label{eq6}
\end{equation}
where the dust cross section $\sigma_{\mathrm{dust}}(\nu)$ is scaled from the reference value at $5500 \ \mathring{\mathrm{A}}$ using an extinction law:
\begin{equation}
    \sigma_{\mathrm{dust}}(\nu) = \sigma_\nu \frac{el(\nu)}{el(c / 5500 \ \mathring{\mathrm{A}})}.
    \label{eq7}
\end{equation}
Here, $el(\nu)$ represents the chosen extinction law.

\begin{figure}
    \centering
    \includegraphics[width=1.0\linewidth]{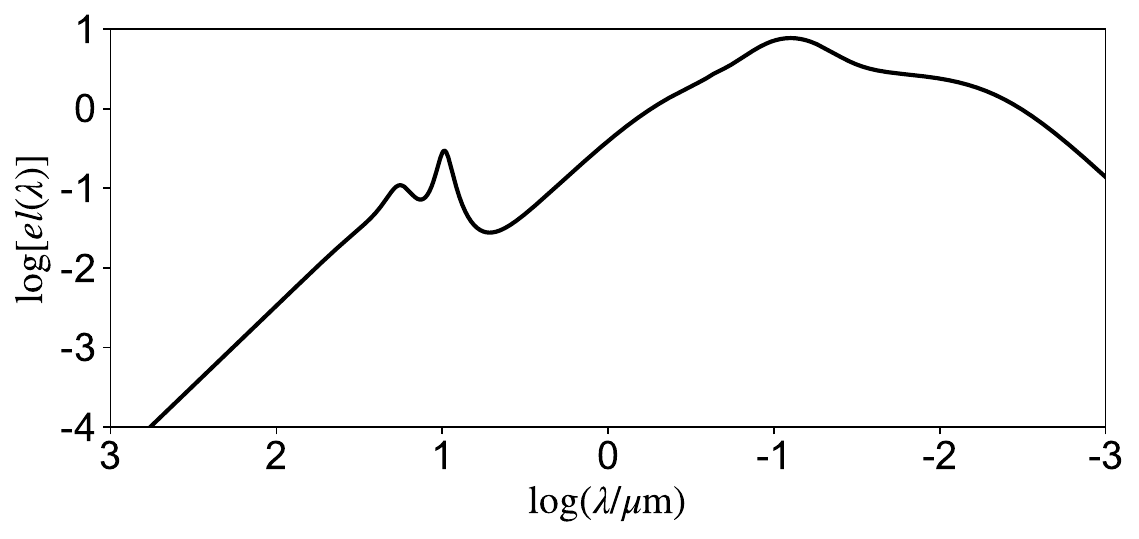}
    \caption{The extinction curve for SMC-type calculated by the model proposed by \cite{1992ApJ...395..130P}. This extinction curve is normalized by $\xi(5500 \ \mathring{\mathrm{A}})=1$.}
    \label{fig3}
\end{figure}

Eq.\ref{eq5} and Eq.\ref{eq6} share the same assumption that the comoving number density of the absorber (IGM/dust) remains constant. Using a Small Magellanic Clouds (SMC)-type extinction law, \cite{2015ApJ...802L..16X} found that this assumption can explain the observed change in the power-law index of quasar spectra due to dust extinction at low redshifts, but a discrepancy arises when applied to high-redshift cases. \cite{2015ApJ...802L..16X} highlighted two potential solutions to this problem. The first approach is to adopt a Milky Way-like extinction curve while retaining the constant comoving number density assumption. The alternative is to adopt a lower comoving number density of dust at high redshift, thereby accounting for dust generation. In this work, we choose the latter method. Based on studies of the infrared background, \cite{2025ApJ...992...65C} modeled the evolution of cosmic dust and provided the best-fit parameters using Bayesian analysis. We adopt their dust evolution model, scaling it to $n\sigma_\nu = 2\times10^{-5} h \ \mathrm{Mpc}^{-1}$ at $z=0$ and using the SMC-like extinction curve given by \cite{1992ApJ...395..130P}, as suggested by \cite{2015ApJ...802L..16X}. Fig.~\ref{fig3} shows the extinction curve normalized by $el(5500 \ \mathring{\mathrm{A}})=1$.

The corresponding evolution of the dust comoving density is
\begin{equation}
    \label{eq8}
    n_d(z) = n_0 \frac{(1+z)^b}{1 + \left[ (1+z) / c_0 \right]^d},
\end{equation}
where $b=1.54$, $c_0=2.83$, and $d=6.55$ are the best-fit parameters. We ignore the contribution of very cold dust, as \cite{2025ApJ...992...65C} found that the best-fit fraction of very cold dust is not significantly greater than zero. Combining Eq.\ref{eq8} with Eq.\ref{eq6} and Eq.\ref{eq7}, the optical depth due to dust is given by:
\begin{equation}
    \label{eq9}
    \tau_{\mathrm{dust}}(\nu, z) = \frac{c n_0 \sigma_\nu}{H_0} \int_0^z \frac{(1+z')^b}{1 + \left[ (1+z') / c_0 \right]^d} \frac{el(\nu(1+z'))}{el(c / 5500 \ \mathring{\mathrm{A}})} \frac{(1+z')^2}{\sqrt{(1+z')^3 \Omega_{\mathrm{m},0} + \Omega_{\Lambda,0}}}  dz',
\end{equation}
with $n_0 \sigma_\nu = 2\times10^{-5} h \ \mathrm{Mpc}^{-1}$. Finally, the received spectrum, after accounting for absorption, is calculated as:
\begin{equation}
    \label{eq10}
        F_{\nu,\mathrm{absorbed}}(\nu, z,\theta,Z) = \frac{1+z}{4\pi D_{\mathrm{L}}^2} L_{\nu,\mathrm{apparent}}(\nu(1+z),\theta) \\
        \times \exp \left[ - \left( \tau_{\mathrm{IGM}}(\nu,z,Z) + \tau_{\mathrm{dust}}(\nu,z) \right) \right].
\end{equation}

\subsection{Radio emission}
The high-speed wind produced in a TDE could interact with the CNM, driving a forward shock into the CNM. This shock accelerates cosmic-ray electrons, which then produce radio emission via synchrotron radiation (\cite{1998ApJ...499..810C, 2013ApJ...772...78B}). In this work, we calculate the radio flare generated by the wind-CNM interaction using the wind properties derived from our simulation.

Following the model presented in \cite{2025ApJ...979..109Z}, we describe the hydrodynamic evolution of the wind using the equations from \cite{2000ApJ...543...90H}:
\begin{align}
    \frac{dr}{dt} &= \beta_w c, \label{eq11} \\
    \frac{dM}{dr} &= \Omega_w r^2 n_{\mathrm{CNM}} m_{\rm p}, \label{eq12} \\
    \frac{d\beta_w}{dM} &= -\frac{\beta_w (1 + \beta_w^2/4)}{m_w + M + (1-\epsilon)(1 + \beta_w^2) M}, \label{eq13}
\end{align}
where $\beta_w = v_w/c$ is the wind velocity normalized by the speed of light, $m_{\rm p}$ is the proton mass, $\Omega_w$ is the wind solid angle, and $\epsilon$ is the radiation efficiency as described in \cite{1999ApJ...520..634D}. The CNM number density $n_{\mathrm{CNM}}$ is modeled with a power-law profile, $n_{\mathrm{CNM}}(r) = A r_{-2}^{-1} \, \mathrm{cm^{-3}}$ (\cite{2025ApJ...979..109Z}), where $r_{-2} = r / (10^{-2} \, \mathrm{pc})$ and $A$ is a constant. We adopt a typical value of $A = 8.5 \times 10^3 \, \mathrm{cm^{-3}}$ based on the fit to AT2019azh presented in \cite{2025ApJ...979..109Z}.

Here, $m_w$ denotes the mass of the wind. Our simulations \cite{2025arXiv251221500S} show that both the mass flux and the kinetic luminosity of the wind peak at approximately $100$ days. For Model M300-9, the peak mass flux is $\sim 2.64 \times 10^{3} \, \dot{M}_{\rm Edd}$, and the peak kinetic luminosity is $\sim 218 \, L_{\rm Edd}$. For Model M300-5, the corresponding peak values are $1.386 \times 10^{3} \, \dot{M}_{\rm Edd}$ and $\sim 135 \, L_{\rm Edd}$, respectively.

We model the radio emission by tracking the wind evolution during its most powerful phase (around \( t \sim 100 \) days), assuming a characteristic duration of \( t_w = 30 \) days, in which we assume the wind properties remain constant and equal to the wind properties calculated at \( t \sim 100 \). The wind is divided into 90 angular bins over \( 0 \le \theta \le \pi/2 \). The mass flux and kinetic luminosity values for each bin, taken from our previous simulation results (\cite{2025arXiv251221500S}), inherently represent the integrated contribution from the full azimuthal range (\( 0 \le \phi \le 2\pi \)) and include a factor of 2 to account for the symmetric counterpart below the mid-plane. For each bin, the mean wind velocity is derived from $\sqrt{2 \, L_{\mathrm{kin,i}} / \dot{M}_{\mathrm{wind,i}}}$, where $L_{\mathrm{kin,i}}$ is the kinetic luminosity and $\dot{M}_{\mathrm{wind,i}}$ is the outflow mass flux of bin $i$ at $t = 100$ days. This provides the initial wind velocity for each bin. The wind mass in each bin is given by $m_w = \dot{M}_{\mathrm{wind,i}} \, t_w$. Using Eqs.~\eqref{eq11}--\eqref{eq13}, we then compute the subsequent velocity evolution of the wind independently for each angular bin.

Having calculated the velocity evolution, we then computed the radio emission following the formalism presented in \cite{2021MNRAS.507.4196M}. The amplified magnetic field strength is given by assuming a fraction $\epsilon_{\mathrm{B}}$ of the post-shock energy is converted into magnetic field energy:
\begin{equation}
\label{eq14}
    B = (8\pi \epsilon_{\mathrm{B}} m_{\mathrm{p}} n_{\mathrm{CNM}} v_w^2)^{1/2},
\end{equation}
where $\epsilon_{\mathrm{B}}$ is set to $0.001$ based on the fit to AT2019azh from \cite{2025ApJ...979..109Z}.

The minimum Lorentz factor $\gamma_{\mathrm{m}}$ of the accelerated electrons is
\begin{equation}
\label{eq15}
    \gamma_{\mathrm{m}} = \max \left[ 2,\; \frac{m_{\mathrm{p}}}{4 m_{\mathrm{e}} c^2} \bar{\epsilon}_{\mathrm{e}} v_{w}^2 \right],
\end{equation}
where $m_{\mathrm{e}}$ is the electron mass, and $\bar{\epsilon}_{\mathrm{e}}$ is defined as $\bar{\epsilon}_{\mathrm{e}} \equiv 4\epsilon_{\mathrm{e}}(p-2)/(p-1)$. Here, $p$ is the power-law index of the electron energy distribution, which we take to be $p = 2.5$, and $\epsilon_{\mathrm{e}}$ is the fraction of shock energy transferred to relativistic electrons. Following \cite{2025ApJ...979..109Z}, we set $\epsilon_{\mathrm{e}} = 0.01$ based on their fit to AT2019azh.

The synchrotron frequency is then given by
\begin{equation}
\label{eq16}
    \nu_{\mathrm{m}} = \gamma_{\mathrm{m}}^2 \frac{e B}{2\pi m_{\mathrm{e}} c},
\end{equation}
where $e$ is the electron charge. A critical velocity is defined as
\begin{equation}
    \label{eq17}
    v_{\mathrm{DN}} = \left( \frac{8 m_{\mathrm{e}}}{m_{\mathrm{p}} \bar{\epsilon}_{\mathrm{e}}} \right)^{1/2}.
\end{equation}
The spectral power per electron at frequency $\nu_{\mathrm{m}}$ is
\begin{equation}
    \label{eq18}
    P_{\nu_{\mathrm{m}}} \simeq \frac{\frac{4}{3} \sigma_{\mathrm{T}} c \gamma_{\mathrm{m}}^2 \frac{B^2}{8\pi}}{\nu_{\mathrm{m}}},
\end{equation}
where $\sigma_{\mathrm{T}}$ is the Thomson cross section. Estimating the number of radiating electrons as $N_{\mathrm{e}} = \Omega_w n_{\mathrm{CNM}} r^3$, the luminosity at $\nu_{\mathrm{m}}$ is
\begin{equation}
    \label{eq19}
    L_{\nu_{\mathrm{m}}} = N_{\mathrm{e}} \, \min\left[ \left( \frac{v_w}{v_{\mathrm{DN}}} \right)^2,\; 1 \right] P_{\nu_{\mathrm{m}}}.
\end{equation}

The synchrotron self-absorption (SSA) frequency is approximately
\begin{equation}
    \label{eq20}
    \nu_{\mathrm{a}} \simeq \left( \frac{(p-1) \pi^{3/2} 3^{\frac{p+1}{2}}}{4} \frac{e n_{\mathrm{CNM}} r \, \min\left[ (v / v_{\mathrm{DN}})^2, 1 \right]}{\gamma_{\mathrm{m}}^5 B} \right) ^{\frac{2}{p+4}}\nu_{\mathrm{m}}.
\end{equation}

In our results, $\nu_{\mathrm{a}} > \nu_{\mathrm{m}}$; therefore, we smooth the spectrum using the prescription from \cite{2002ApJ...568..820G}:
\begin{equation}
    \label{eq21}
    L_{\nu} = L_{\nu_{\mathrm{m}}} \left[ \left( \frac{\nu}{\nu_{\mathrm{m}}} \right)^2 \exp\left( -s_4 (\nu/\nu_{\mathrm{m}})^{2/3} \right) + \left( \frac{\nu}{\nu_{\mathrm{m}}} \right)^{5/2} \right] \times \left[ 1 + \left( \frac{\nu}{\nu_{\mathrm{a}}} \right)^{s_5 (\beta_2 - \beta_3)} \right]^{-1/s_5},
\end{equation}
where $\beta_2 = 5/2$, $\beta_3 = (1 - p)/2$, $s_4 = 3.63p - 1.6$, and $s_5 = 1.25 - 0.18p$, following \cite{2025ApJ...979..109Z}. The final radio spectrum at any given time is obtained by summing the contributions from all individual bins.

\section{RESULTS}
Since the results of two models are similar to each other, this section is devoted to the results derived from model M300-9. A comprehensive set of results for model M300-5 can be found in the Appendix.

\subsection{Blackbody emission}

Fig.~\ref{fig4} presents the temporal evolution of the apparent specific luminosity spectrum along five distinct lines of sight. In general, the specific luminosities observed from different viewing angles are similar at a given snapshot in the low-energy bands, such as the infrared and optical wavebands. However, a clear viewing-angle dependence emerges in the extreme ultraviolet (EUV) and X-ray wavebands. This trend is consistent with the simulation results reported in the non-jetted model of \cite{Curd&Narayan2019}, where the emission properties were derived by post-processing their simulation data using the radiative transfer code HEROIC. This angular dependence is caused by the structure of the photosphere.
\begin{figure}
    \centering
    \includegraphics[width=0.8\linewidth]{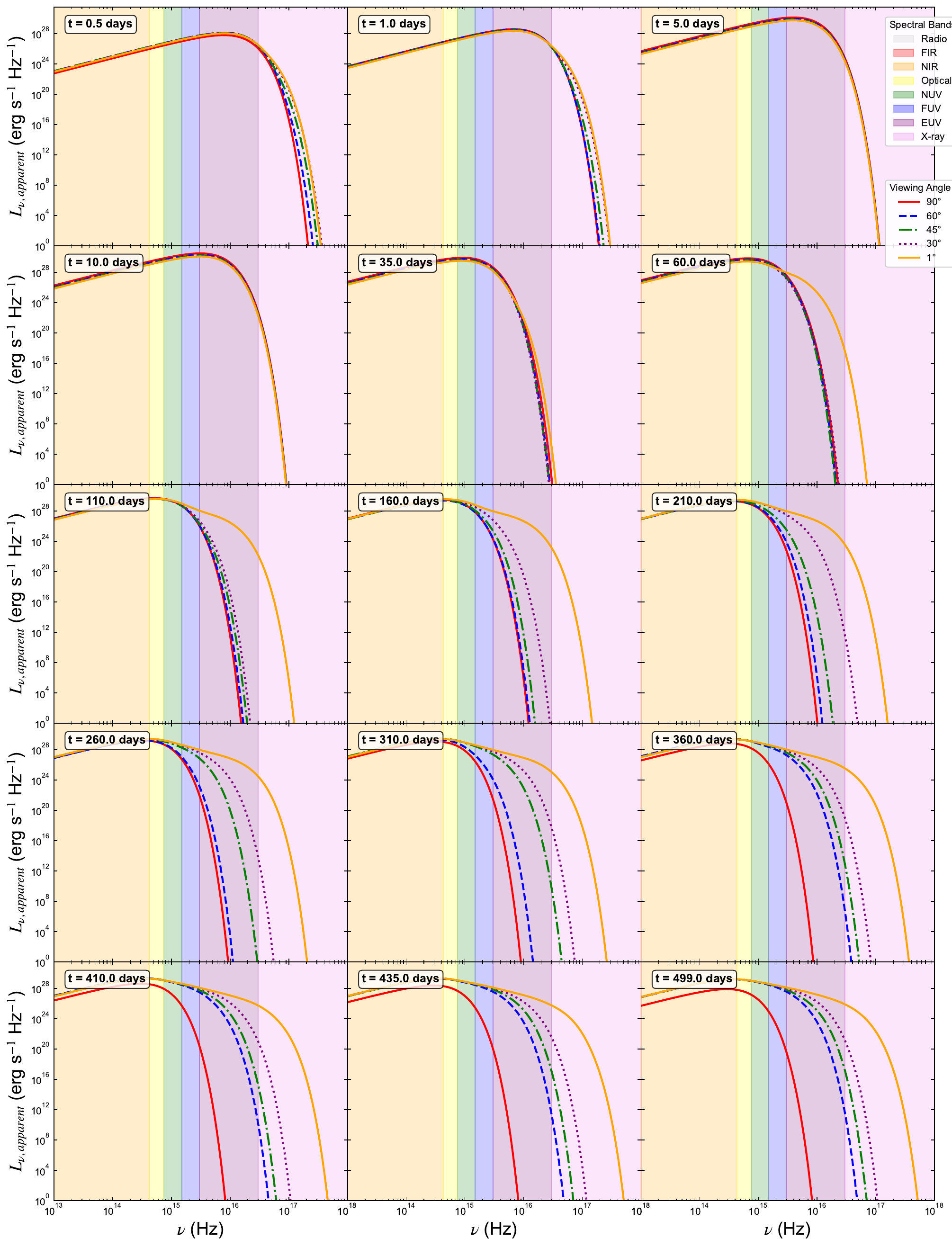}
    \caption{Temporal evolution of the apparent specific luminosity for model M300-9, shown for 15 discrete snapshots from $t = 0.5$ to $499.0\,\mathrm{days}$. The coloured background bands indicate different spectral wavebands: radio (gray), far-infrared (FIR; red), near-infrared (NIR; orange), optical (yellow), near-ultraviolet (NUV; green), far-ultraviolet (FUV; blue), extreme ultraviolet (EUV; purple), and X-ray (pink). Spectra are shown for different viewing angles: $\theta = 1^\circ$ (solid orange line), $30^\circ$ (dotted purple line), $45^\circ$ (dash-dotted green line), $60^\circ$ (dashed blue line), and $90^\circ$ (solid red line).}
    \label{fig4}
\end{figure}

In general, the radiation temperature decreases with an increasing radius due to the adiabatic expansion of the trapped photons. Consequently, high-energy photons are predominantly produced in the inner regions of the system (at smaller radii). In normal Pop I star TDEs, an optically thin funnel forms along the rotational axis, which allows X-rays generated in the central region to escape directly (\cite{BU22, Dai2018, 2019ApJ...880...67J, Thomsen2022, Curd&Narayan2019}). In contrast, photons originating from the extended outer photosphere are emitted at lower temperatures and thus carry lower energies. Since the outer regions of the extended photosphere rarely obscure each other, the specific luminosities observed from different lines of sight are similar in low-energy bands. However, when viewed from large angles (i.e., large $\theta$), high-energy photons produced within the funnel can be obscured by the outer photosphere, leading to significant attenuation. This results in a noticeable decrease in the observed high-energy flux as the viewing angle increases from alignment with the rotational axis ($\theta \sim 0^\circ$) toward the mid-plane ($\theta \sim 90^\circ$).

However, such a funnel region can be diminished under extremely high accretion rates (\cite{2016MNRAS.456.3929S}), a phenomenon we also find in our simulation due to the extreme accretion rate of Pop III star TDEs. In our simulation, the photosphere is initially elongated perpendicular to the mid-plane. At early times, the location of the photosphere along the rotational axis is larger than that in the equatorial plane and there are no funnel regions produced (see the `$t=10.0\,\mathrm{days}$' panel in Fig.~\ref{fig1}). The photosphere radius along the axis reaches its maximum of $3.4\times10^4\,R_{\rm s}$ (where $R_{\rm s}$ is the Schwarzschild radius) at $t=21.8\,\mathrm{days}$. As the accretion rate decreases with time, the photosphere along the rotational axis gradually recedes toward the black hole, and a funnel region forms during this process. By $t=449\,\mathrm{days}$, the photosphere at $\theta=0^\circ$ has receded to our simulation's inner boundary at $r=2R_{\rm s}$, and the overall photospheric shape becomes elongated along the mid-plane at late times (see the `$t=499.0\,\mathrm{days}$' panel in Fig.~\ref{fig1}).

The temporal evolution of the specific luminosity shown in Fig.~\ref{fig4} can therefore be explained by the structural evolution of the photosphere. At early times, the photosphere has not yet expanded to a large radius, maintaining a relatively high temperature capable of emitting high-energy photons. Since the mutual obscuration is weak during this period, the dependence on the line of sight is insignificant (see the `$t=0.5\,\mathrm{days}$' panel in Fig.~\ref{fig4}). The spectrum peaks in the EUV band with a specific luminosity of $1.0\times10^{28}~\mathrm{erg~s^{-1}~Hz^{-1}}$ at this stage.

As the system evolves, the photosphere expands to larger radii, and its radiation temperature decreases. During the early expansion phase (from the `$t=0.5\,\mathrm{days}$' to the `$t=10.0\,\mathrm{days}$' panel in Fig.~\ref{fig4}), the increase in photospheric surface area dominates, leading to a rise in specific luminosity from $1.0\times10^{28}$ to $3.0\times10^{30}~\mathrm{erg~s^{-1}~Hz^{-1}}$, while the spectral peak shifts to lower frequencies due to the dropping temperature.

Once the photosphere expands further and the direct influence of the accretion region weakens, it enters a phase of nearly adiabatic expansion. In our simulation, the radial density profile of the wind follows $\rho \propto r^{-2}$, as expected from mass conservation for a steady wind. For adiabatic expansion, the temperature scales as $T \propto \rho^{1/3}$. Consequently, the luminosity evolves as $L \propto r^{2} T^{4} \propto r^{2} \rho^{4/3} \propto r^{-2/3}$. This scaling implies that, for a steadily expanding adiabatic wind, the decrease in temperature outweighs the increase in emitting area. Therefore, after $t \approx 10.0\,\mathrm{days}$, the specific luminosity begins to decrease, as seen from the `$t=10.0\,\mathrm{days}$' to the `$t=499.0\,\mathrm{days}$' panel in Fig.~\ref{fig4}.

Subsequently, as the photosphere along the rotational axis begins to recede, the funnel forms and penetrates deeper towards the black hole, thus its radiation temperature increases over time. Consequently, the specific luminosity along $\theta=1^\circ$ shows more high-energy photons, and the luminosity in the high energy bands increases with time. However, due to obscuration by the outer photosphere, this high-energy emission remains invisible at larger viewing angles ($\theta$). Thus, a significant viewing-angle dependence emerges initially only between $\theta=1^\circ$ and other angles at this time (see the `$t=60.0\,\mathrm{days}$' panel in Fig.~\ref{fig4}).

This viewing-angle dependence extends to other angles at later times as the photosphere morphology evolves. During the transition from a vertically elongated structure (see the `$t=60.0\,\mathrm{days}$' panel in Fig.~\ref{fig1}) to a horizontally elongated one (see the `$t=499.0\,\mathrm{days}$' panel in Fig.~\ref{fig1}), the opening angle of the funnel increases. This allows the inner, hotter parts of the photosphere to become visible to observers at progressively larger angles: first $\theta=30^\circ$, then $\theta=45^\circ$, and finally $\theta=60^\circ$. Nevertheless, the innermost part of the photosphere remains observable only at a small $\theta$ due to its small opening angle, preserving differences between these lines of sight. The viewing-angle dependence thus strengthens in high-energy bands at late times (from the `$t=60.0\,\mathrm{days}$' to `$t=499.0\,\mathrm{days}$' panels in Fig.~\ref{fig4}).

Over long-term evolution, as the photosphere cools, the spectral peak shifts from the NUV ( see the ‘t = 35.0 days’ panel in Fig.~\ref{fig4}) to the NIR/Optical ( see the ‘t = 499.0 days’ panel in Fig.~\ref{fig4}). Concurrently, the peak specific luminosity decreases from $8.1 \times 10^{29} \ \mathrm{erg \ s^{-1} \ Hz^{-1}}$ at $t = 35.0 \ \mathrm{days}$ to $1.6 \times 10^{29} \ \mathrm{erg \ s^{-1} \ Hz^{-1}}$ at $t = 499.0 \ \mathrm{days}$. During this period, these outer photosphere elements are visible from most lines of sight, except along the mid-plane. As the photosphere becomes horizontally elongated, many surface elements become invisible at $\theta=90^\circ$ (see the `$t=499.0\,\mathrm{days}$' panel in Fig.~\ref{fig1}), causing the optical/IR luminosity at $\theta=90^\circ$ to become lower (the peak specific luminosity is $7.7\times10^{27}~\mathrm{erg~s^{-1}~Hz^{-1}}$ along $\theta=90^\circ$) than that from other viewing angles (see the `$t=499.0\,\mathrm{days}$' panel in Fig.~\ref{fig4}).

In our previous work (\cite{2025arXiv251221500S}), we simply defined t = 449.0 days—when the photosphere along the axis recedes to the BH surface—as the time when X-rays from the disk could leak freely. In reality, the blackbody emission from the photosphere always has a high-energy X-ray tail. This component always exists and grows stronger as the photospheric temperature rises during funnel expansion. Consequently, the X-ray flux increases continuously (as shown in Fig. 4), becoming prominent even before 449.0 days.

\begin{figure}
    \centering
    \includegraphics[width=0.8\linewidth]{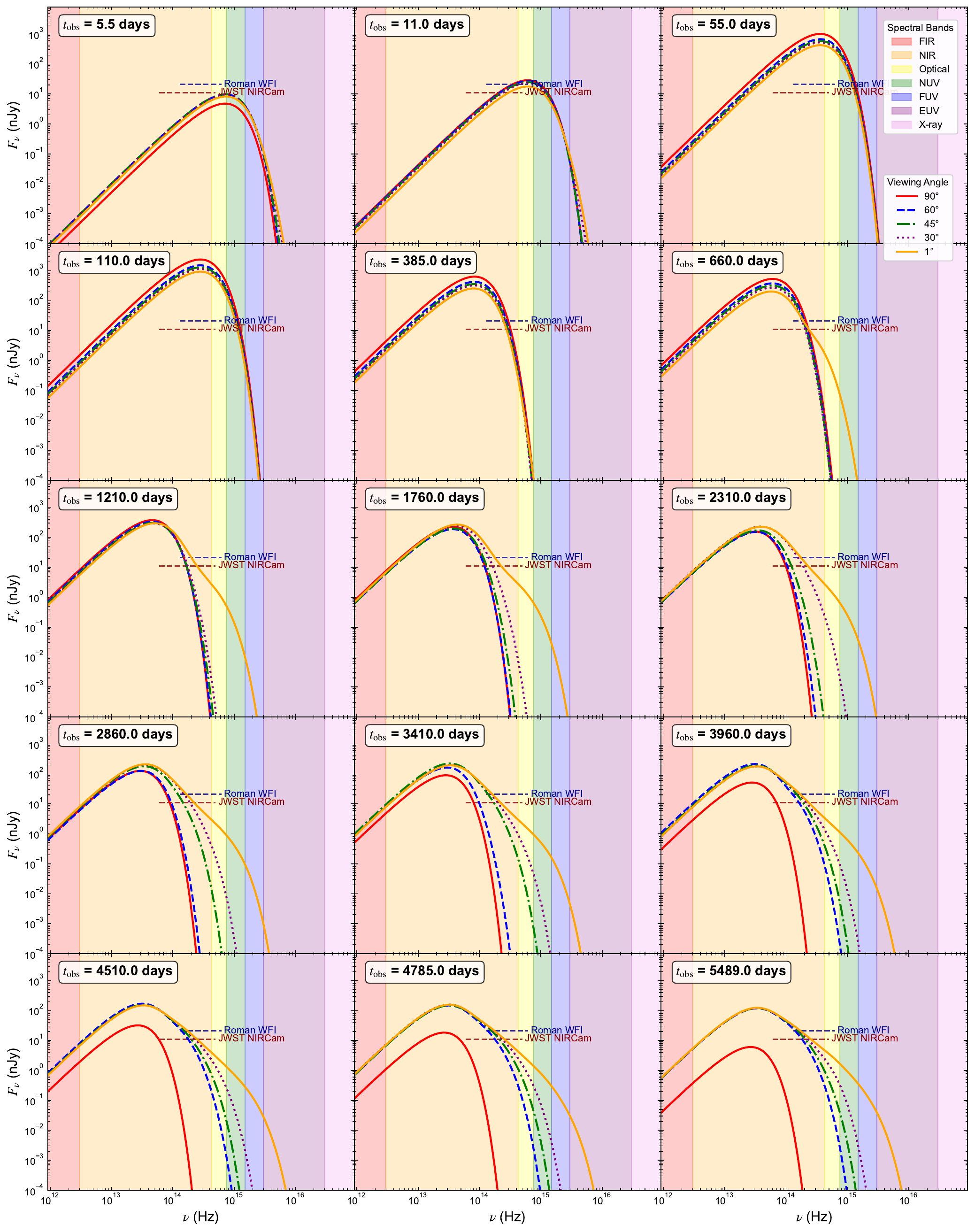}
    \caption{Temporal evolution of the flux density for model M300-9, displayed at 15 discrete snapshots from $t_{\rm obs} = 5.5$ to $5489.0\,\mathrm{days}$, where $t_{\rm obs}$ is the observer frame time. The calculation assumes a cosmological redshift of $z = 10$ and does not include extinction effects. The horizontal red and blue dashed lines indicate the detection limits of JWST and the Roman Space Telescope, respectively. The color bands (denoting spectral wavebands) and line styles (representing different viewing angles) follow the same convention as in Fig.~\ref{fig4}.}
    \label{fig5}
\end{figure}
\begin{figure}
    \centering
    \includegraphics[width=0.9\linewidth]{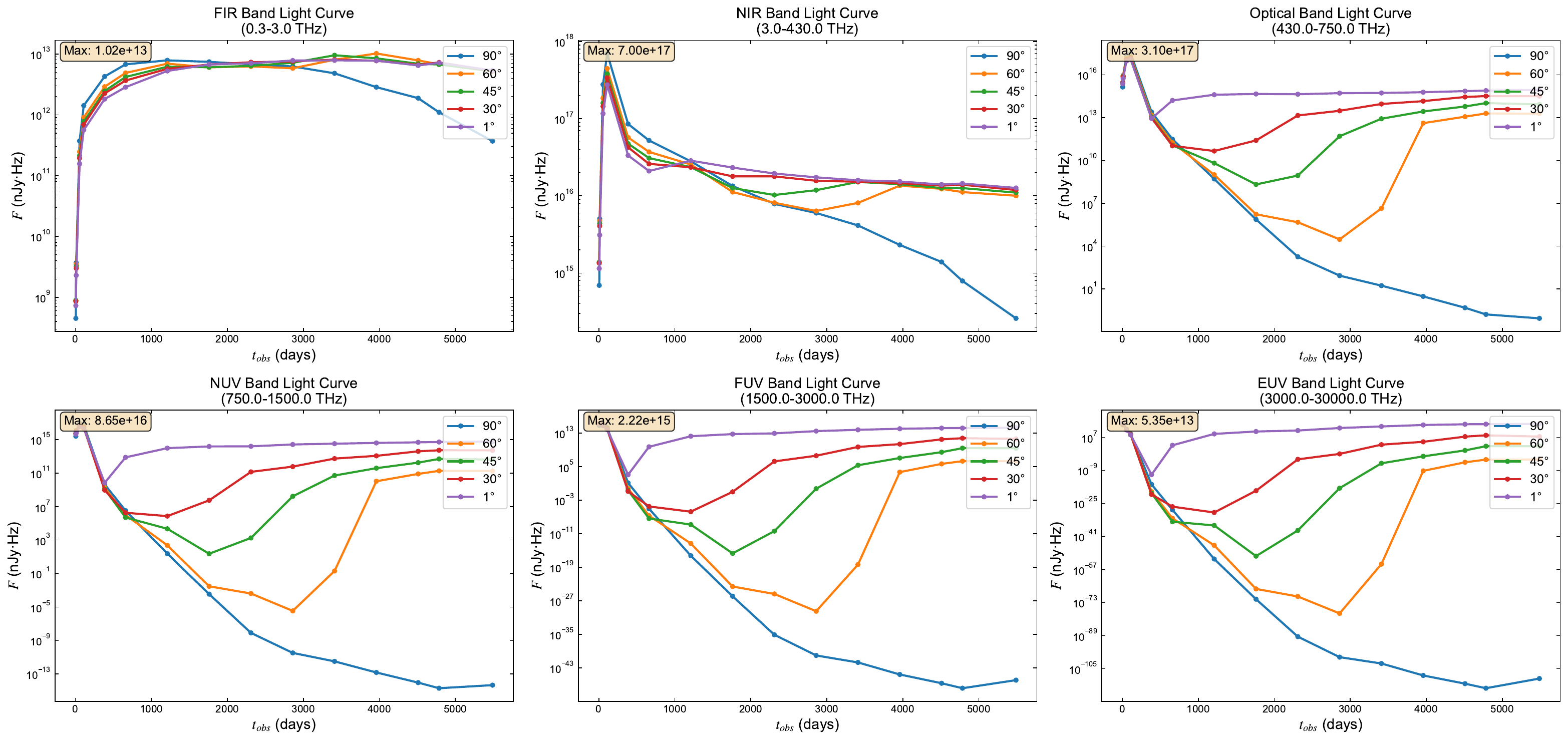}
    \caption{Received light curves in different wavebands for model M300-9, assuming a cosmological redshift of $z = 10$ and no extinction. The six panels show the integrated flux as a function of time in the following bands (from top-left to bottom-right): far-infrared (FIR), near-infrared (NIR), optical, near-ultraviolet (NUV), far-ultraviolet (FUV), and extreme ultraviolet (EUV). In each panel, the light curves are plotted for five different viewing angles: $\theta = 1^\circ$ (purple), $\theta = 30^\circ$ (red), $\theta = 45^\circ$ (green), $\theta = 60^\circ$ (orange), and $\theta = 90^\circ$ (blue).}
    \label{fig6}
\end{figure}

 Fig.~\ref{fig5} presents the observed flux density without accounting for absorption, assuming a source redshift of $z=10$. Due to cosmological time dilation, the observed timescale is stretched by a factor of 11. A significant portion of the radiative energy is shifted into the IR band, and a clear viewing-angle dependence emerges in the optical/UV bands. In Fig.~\ref{fig5}, we also plot the detection limits of the NIRCam on JWST ($\sim 11\,\mathrm{nJy}$, corresponding to the F150W filter, covering 600–5000 nm) and the WFI on Roman ($\sim 20.9\,\mathrm{nJy}$, corresponding to the F106 filter, covering 480–2300 nm). It can be seen that the Pop III star TDE has the potential to be detected by JWST throughout most of its evolution (except for views along $\theta=90^\circ$, which become undetectable at late times due to the change in photosphere morphology). For Roman's WFI at a relatively later time, however, detection may only be feasible along sight lines with small $\theta$ .

Figure~\ref{fig6} displays the light curves across various wavebands for the source, calculated without accounting for absorption and assuming a redshift of $z=10$. The light curves generally show an initial rise due to the expansion of the photosphere, followed by a decline driven by the adiabatic cooling process described above. However, the light curve in the FIR band continues to increase over an extended period. This behavior arises because the blackbody intensity in this low-energy regime follows the Rayleigh-Jeans approximation $B_{\nu} \simeq (2kT/c^2)\nu \propto T$. The integrated luminosity in the band thus scales as $L \propto r^2 \int B_{\nu} d\nu \propto r^2 T \propto r^2 \rho^{1/3} \propto r^{4/3}$. Although the overall bolometric luminosity decreases as $L \propto T^4$ during adiabatic expansion, the luminosity in low-frequency bands can still increase with the growing photospheric radius.

In contrast, the light curve in the near-infrared (NIR) band—which originates from the higher-energy optical band in the rest frame and is therefore more sensitive to temperature variations—exhibits a decline after the initial rise, followed by a slight rebrightening due to the exposure of the hot inner region. The viewing-angle dependence in these bands is manifested primarily in the different times at which the flux begins to increase for different lines of sight. However, as noted earlier, the amplitude of the flux variation remains similar across most viewing angles (except for $\theta = 90^\circ$ at late times) in the NIR band.

For the higher-energy bands (Optical to EUV), which correspond to even higher-energy photons (NUV to X-ray) in the rest frame, the light curves exhibit a characteristic two-phase evolution: an initial rapid decline due to the overall temperature decrease, followed by a subsequent rise caused by the exposure of the inner hot region (again, except for $\theta = 90^\circ$, where the rising phase is absent due to obscuration). The angular dependence is significant in these wavebands. The light curves in these high-energy bands shown in Fig.~\ref{fig6} appear similar in shape because the high-energy photons primarily originate from the same physical region—the hot funnel—and their temporal evolution is governed by the shared process of the funnel's exposure to the line of sight.
\begin{figure}
    \centering
    \includegraphics[width=0.8\linewidth]{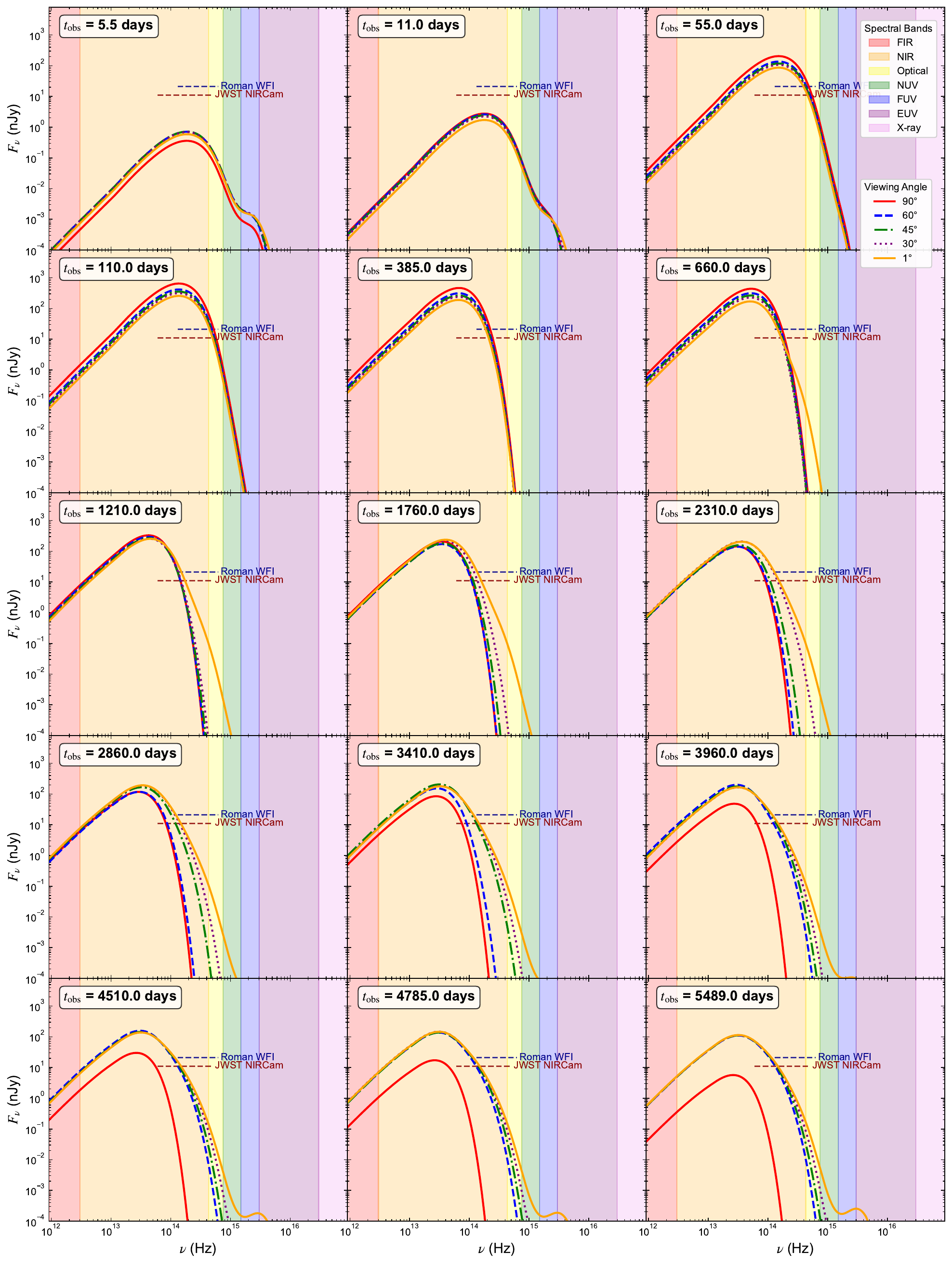}
    \caption{Received flux density for model M300-9, calculated using the extinction model described in this work. The data presentation and viewing angles follow the same convention as in Fig.~\ref{fig5}.}
    \label{fig7}
\end{figure}
\begin{figure}
    \centering
    \includegraphics[width=0.9\linewidth]{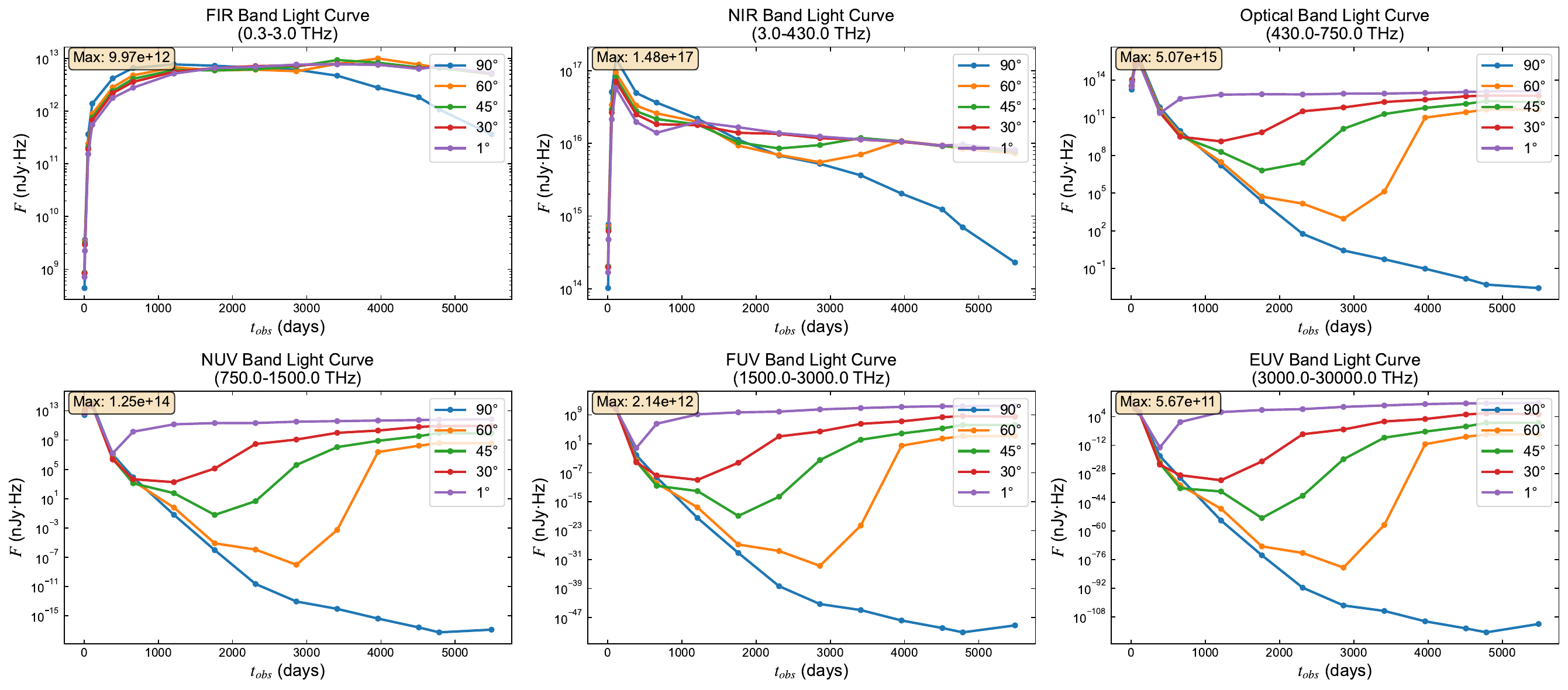}
    \caption{Received light curves in different wavebands for model M300-9, calculated using the extinction model described in this work. The data presentation and viewing angles follow the same convention as in Fig.~\ref{fig6}.}
    \label{fig8}
\end{figure}

After accounting for the extinction of both the IGM and dust using the model described in Section~\ref{sectionextinction}, Fig.~\ref{fig7} shows the resulting flux density. Due to the significant extinction effects—dust primarily affecting the optical band and gas affecting the EUV and X-ray bands—the flux in these wavebands is reduced to very low levels. As a result, given the sensitivity limits of our observations, the viewing-angle dependence—which is evident in the unabsorbed spectra (Fig.~\ref{fig5})—becomes nearly undetectable in these bands after extinction is applied. Detecting this dependence would require substantially greater observational sensitivity than is currently available. A slight degree of viewing-angle dependence briefly reappears during the phase when the inner region begins to be exposed (see the second and third rows of panels in Fig.~\ref{fig7}). However, at late times, this dependence vanishes entirely, making the spectra appear virtually identical across almost all viewing angles. Fortunately, even when extinction is considered, Pop III star TDEs remain detectable by both JWST and Roman, except for observations along $\theta = 90^\circ$ at late times for the reasons stated earlier.

Figure~\ref{fig8} presents the light curves after accounting for extinction. While the overall temporal evolution remains similar to that shown in Fig.~\ref{fig6}, the flux levels are significantly reduced. The reduction is most substantial in the high-energy bands; for instance, the peak flux in the EUV band drops by approximately two orders of magnitude. Even in the infrared bands, where the assumed dust extinction is relatively weak, a slight decrease in flux is still observed. This can be attributed to the redshift effect: during propagation, a low-energy photon that is eventually observed in the infrared may, at some intermediate redshift, have had its frequency blueshifted to a value where dust extinction is significant, thus experiencing attenuation along its path.

\subsection{Radio emission}
\label{subsection3.2}
\begin{figure}
    \centering
    \includegraphics[width=0.9\linewidth]{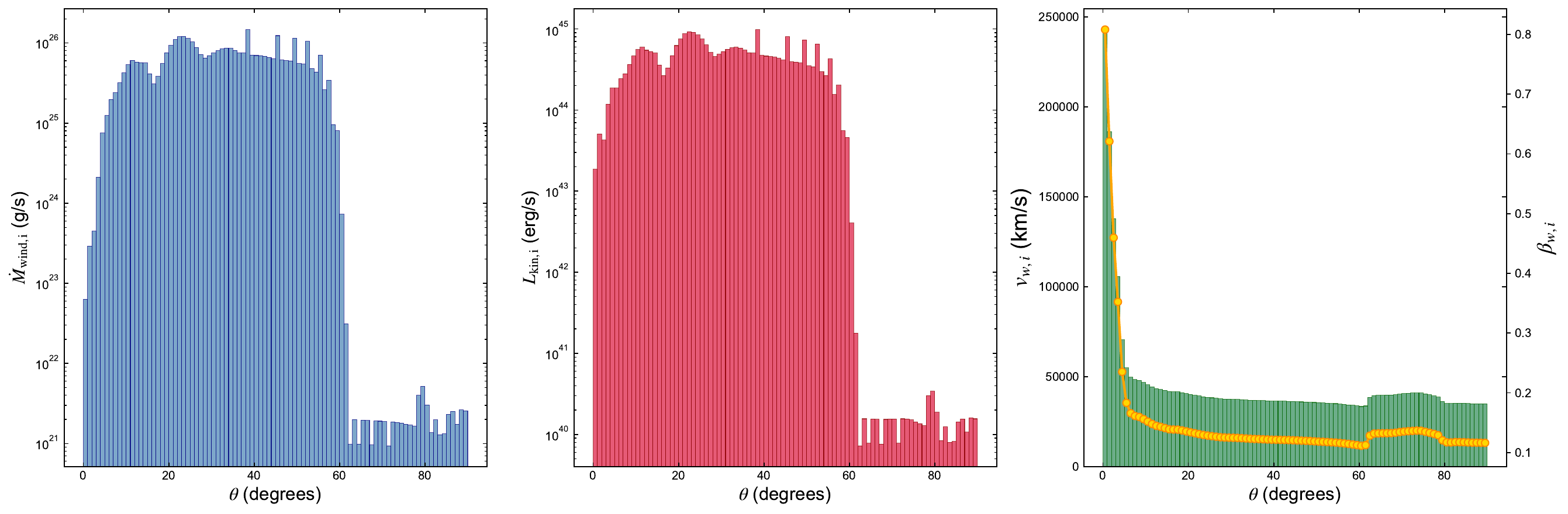}
    \caption{Wind properties across 90 angular bins at $r=10^5R_{\rm s}$ for model M300-9 at $t = 100.0$ days. The left panel shows the mass flux $\dot{M}_{\mathrm{wind},i}$ in each bin. The middle panel displays the corresponding kinetic luminosity $L_{\mathrm{kin},i}$. The right panel presents the wind velocity $v_{\mathrm{w},i}$ (green bins), with the yellow circles and lines indicating the normalized velocity $\beta_{\mathrm{w},i} = v_{\mathrm{w},i}/c$.}
    \label{fig9}
\end{figure}
\begin{figure}
    \centering
    \includegraphics[width=0.9\linewidth]{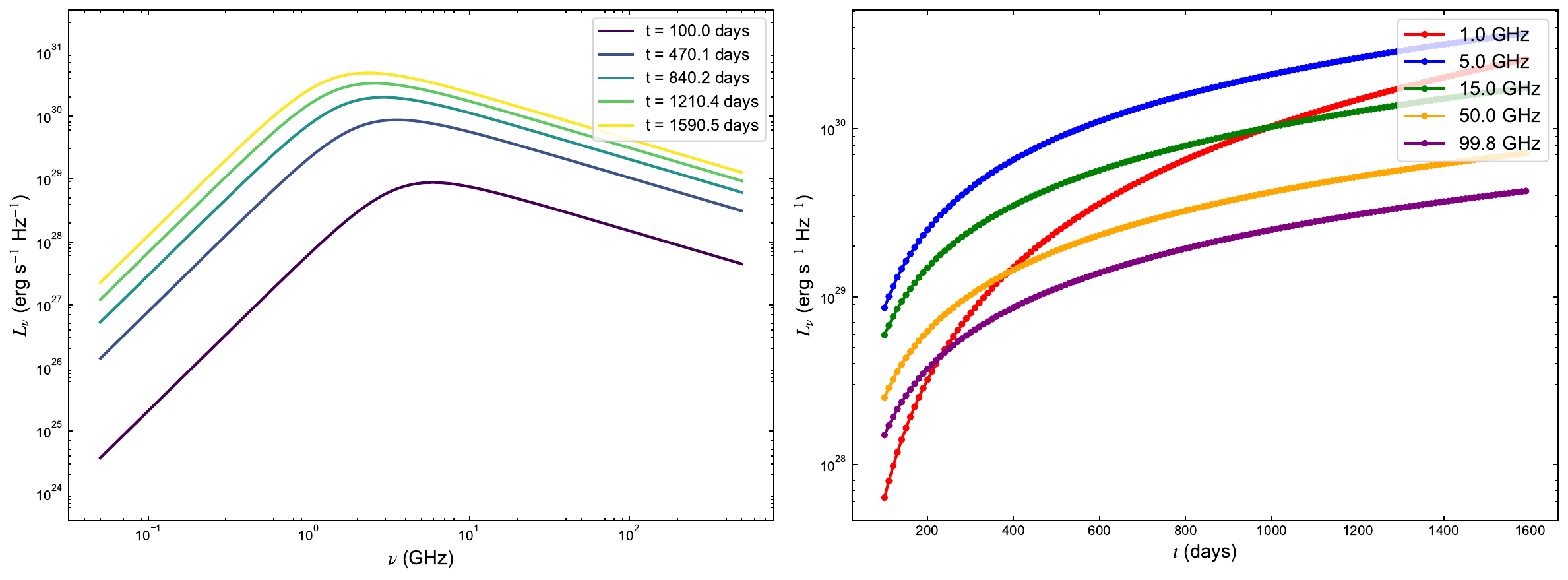}
    \caption{Radio spectral evolution for model M300-9. The left panel shows the spectra at five different times: $t = 100.0\,\mathrm{days}$ (purple), $t = 470.1\,\mathrm{days}$ (dark blue), $t = 840.2\,\mathrm{days}$ (blue), $t = 1210.4\,\mathrm{days}$ (green), and $t = 1590.5\,\mathrm{days}$ (yellow). The right panel shows the temporal evolution of the specific luminosity at five selected frequencies in the rest frame: 1.0 GHz (red), 5.0 GHz (blue), 15.0 GHz (green), 50.0 GHz (orange), and 99.8 GHz (purple).}
    \label{fig10}
\end{figure}
In our previous numerical study, we investigated the properties of the wind launched from the accretion system formed during a Population III star TDE. The wind properties adopted here for the calculation of the radio emission are taken directly from the results of that earlier simulation  \cite{2025arXiv251221500S}.  Figure~\ref{fig9} shows the variation of the wind mass flux with the polar angle $\theta$ across the 90 angular bins. The mass flux initially increases with $\theta$, and then decreases significantly for $\theta \gtrsim 60^\circ$. This behavior occurs because the wind velocity is generally lower at larger $\theta$ in the inner$ r\le100R_{\rm s}$ region (see the Fig.4 of our previous work \cite{2025arXiv251221500S}). Since the wind in the outer region is all comes from the inner region where debris injected, lower inner radial velocity at larger $\theta$ means that the majority of the wind travels along these angles could hardly travels to a larger distance by t=100 days. As a result, at t=100.0 days we calculated here, the wind at $\theta \gtrsim 60^{\circ}$ have not reached the outer boudary of our simulation. There are only gas expands from the higher latitude winds, or the backgrounds gas driven by the shock produced by the higher latitude winds. These gas also has high velocity, but the mass of them are much smaller than the majority of wind in the high latitude, thus we could see a mass flux drop at $\theta \gtrsim 60^{\circ}$, while the velocity could retain a high value as shown in Fig.~\ref{fig9}.

For each angular bin, the radio emission is calculated using the formalism given by Equations~(\ref{eq11})--(\ref{eq21}). The contributions from all bins are then summed to obtain the total radio emission. The resulting intrinsic spectrum and the corresponding light curve in the rest frame are shown in Figure~\ref{fig10}. We note that since the wind properties are calculated at $t=100.0\rm\ days$, the initial snapshot of radio calculation $t_{\rm cal} = 0.0\rm\ days$ (the first panel of Fig.~\ref{fig10}) refers to the rest-frame time $t=100.0\rm\ days$.

The simulation results reveal that over an evolutionary timescale of $1490.5\,\mathrm{days}$, the radio luminosity at all frequencies continues to increase, while the peak frequency—determined by the SSA frequency $\nu_{\mathrm{a}}$ in our $\nu_{\mathrm{m}} < \nu_{\mathrm{a}}$ regime—continuously decreases. This behavior is a direct consequence of the extremely powerful winds generated in Pop III star TDEs. As reported in our previous work \cite{2025arXiv251221500S}, more than $65\%$ of the fallback debris is unbound to form a wind (with a fraction of $\sim 75\%$ for model M300-5). Given a total fallback mass of $\sim 150\,M_{\odot}$ for a $300\,M_{\odot}$ star TDE, the resulting wind mass is enormous. During the $30$-day period considered here, the total ejected mass reaches $\sim 4.83\,M_{\odot}$. This extreme mass makes the wind resistant to deceleration as it interacts with the CNM. At the end of the calculation at $t = 1490.5\,\mathrm{days}$, the ratio of the total swept-up CNM mass to the wind mass is only $\sim 0.0395$, indicating negligible deceleration.
\begin{figure}
    \centering
    \includegraphics[width=0.9\linewidth]{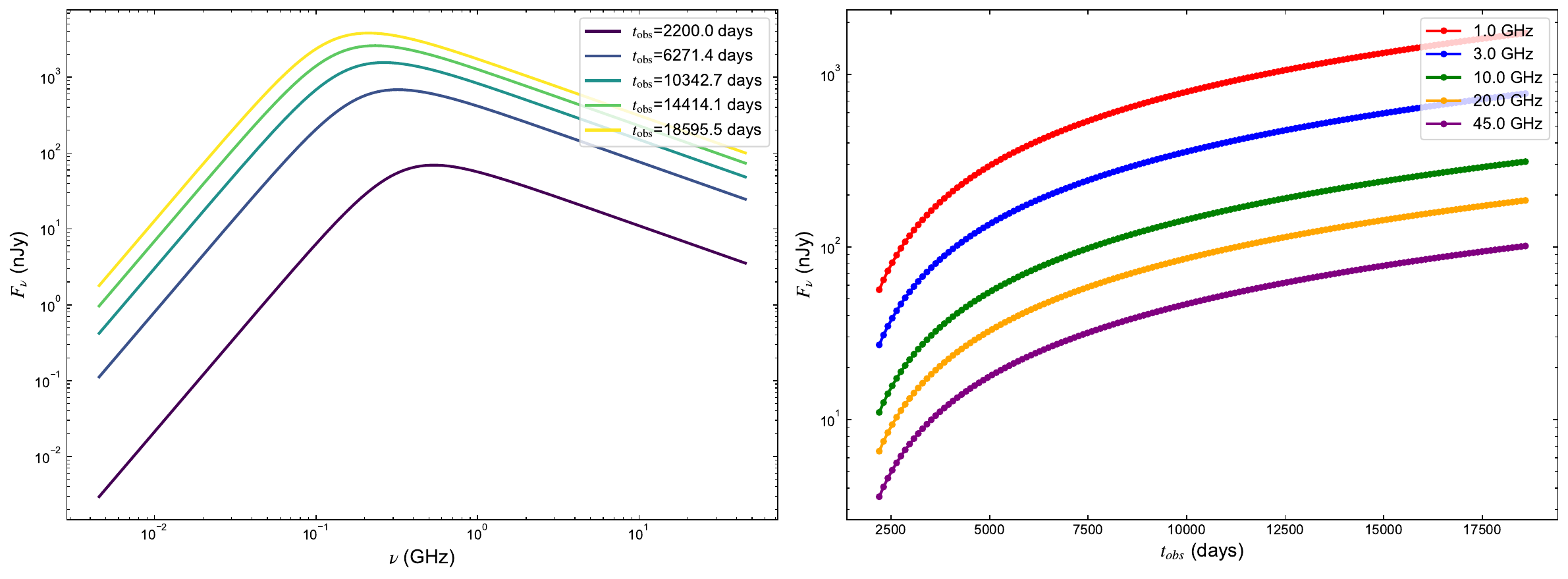}
    \caption{Observed radio flux density for model M300-9 after applying a cosmological redshift of $z=10$. The left panel shows the spectra at five epochs in the observer frame: $t_{\rm obs} = 2200.0\,\mathrm{days}$ (purple), $t_{\rm obs} = 6271.4\,\mathrm{days}$ (dark blue), $t_{\rm obs} = 10342.7\,\mathrm{days}$ (blue), $t_{\rm obs} = 14414.1\,\mathrm{days}$ (green), and $t_{\rm obs} = 18595.5\,\mathrm{days}$ (yellow). The right panel presents the corresponding light curves at five observed frequencies: $1.0\,\mathrm{GHz}$ (red), $3.0\,\mathrm{GHz}$ (blue), $10.0\,\mathrm{GHz}$ (green), $20.0\,\mathrm{GHz}$ (orange), and $45.0\,\mathrm{GHz}$ (purple).}
    \label{fig11}
\end{figure}
Since the wind velocity in most angular bins remains nearly constant, the minimum Lorentz factor $\gamma_{\mathrm{m}}$ of the accelerated electrons shows little evolution. According to Equations~(\ref{eq14}) and (\ref{eq16}), the amplified magnetic field strength scales as $B \propto n_{\mathrm{CNM}}^{1/2}$ and the synchrotron frequency scales as $\nu_{\mathrm{m}} \propto B \propto n_{\mathrm{CNM}}^{1/2}$. The spectral power per electron at $\nu_{\mathrm{m}}$ is given by $P_{\nu_{\mathrm{m}}} \propto B^2 / \nu_{\mathrm{m}} \propto n_{\mathrm{CNM}}^{1/2}$ (Equation~\ref{eq18}). The number of radiating electrons scales as $N_{\mathrm{e}} \propto n_{\mathrm{CNM}} r^3$, leading to a luminosity at $\nu_{\mathrm{m}}$ of $L_{\nu_{\mathrm{m}}} \propto N_{\mathrm{e}} P_{\nu_{\mathrm{m}}} \propto n_{\mathrm{CNM}}^{3/2} r^3$. For the assumed CNM density profile $n_{\mathrm{CNM}}(r) = A r_{-2}^{-1} \, \mathrm{cm^{-3}}$, this results in $L_{\nu_{\mathrm{m}}} \propto r^{3/2}$. Thus, the continuous increase in luminosity is driven by the growing number of radiating electrons as the wind expands, despite the declining CNM density.

On the other hand, the SSA frequency evolves as $\nu_{\mathrm{a}} \propto (n_{\mathrm{CNM}} r / B)^{2/(p+4)} \nu_{\mathrm{m}} \propto \nu_{\mathrm{m}} / B^{2/(p+4)}$ (Equation~\ref{eq20}). For $p = 2.5$, this simplifies to $\nu_{\mathrm{a}} \propto n_{\mathrm{CNM}}^{0.69} \propto r^{-0.69}$. The decrease in CNM density with radius reduces the magnetic field strength, which in turn causes the peak SSA frequency to shift to lower values over time. Consequently, the peak specific luminosity increases from $8.79 \times 10^{28}~\mathrm{erg~s^{-1}~Hz^{-1}}$ at the beginning of the calculation to $4.84 \times 10^{30}~\mathrm{erg~s^{-1}~Hz^{-1}}$ at $t = 1490.5\,\mathrm{days}$, while the peak frequency decreases from $5.9\,\mathrm{GHz}$ to $2.3\,\mathrm{GHz}$.

Finally, Fig.~\ref{fig11} presents the received spectrum and light curve after accounting for a cosmological redshift of $z=10$. The observed radio flux shows a continuous increase over a timescale of $10^4$ days. At late times, the flux exceeds $100\,\mathrm{nJy}$ and can reach levels above $10^3\,\mathrm{nJy}$ at an observed frequency of $\nu_{\mathrm{obs}} = 1.0\,\mathrm{GHz}$. This indicates that Pop~III star TDEs are also promising sources for detection in the radio wavebands, producing an unusually long-lived and continuously rising radio flare with a duration exceeding $10^4$ days. We note that the calculation of radio emission presented here is based on a specific assumed density profile for the circumnuclear medium (CNM). A detailed exploration of how variations in the CNM model affect our results is provided in Section~\ref{Appendix.B}.

\section{CONCLUSION AND DISCUSSION}
In this work, we have calculated the emission properties of a Pop III star TDE system based on the simulation results of \cite{2025arXiv251221500S}. This includes the blackbody emission and the radio flare produced by the interaction between the wind and CNM. To derive observable predictions, we applied an extinction model that accounts for absorption by the WHIM and extinction by dust following an SMC-like law. The dust evolution history is adopted from \cite{2025ApJ...992...65C}, and the model is normalized using the effective cross section from \cite{2015ApJ...802L..16X}. Finally, we have calculated the received spectrum assuming a cosmological redshift of $z=10$.

In the rest frame, we find that the blackbody emission initially peaks in the EUV band, with a specific luminosity of $1.0 \times 10^{28}~\mathrm{erg~s^{-1}~Hz^{-1}}$ at $t = 0.5~\mathrm{days}$ increasing to  $3.0 \times 10^{30}~\mathrm{erg~s^{-1}~Hz^{-1}}$ at $t = 10.0~\mathrm{days}$. The rapid expansion of the photosphere causes its temperature to drop quickly at early times, shifting the peak frequency to the NUV band and reducing the specific luminosity to $6.7 \times 10^{29}~\mathrm{erg~s^{-1}~Hz^{-1}}$ at $t = 60.0~\mathrm{days}$. Subsequently, the temperature continues to decline due to a combination of radiative cooling and decreasing heating from the falling accretion rate. Consequently, the spectral peak shifts further into the NIR band, and at the end of the simulation ($t = 500.0~\mathrm{days}$), the specific luminosity has decreased to $1.6 \times 10^{29}~\mathrm{erg~s^{-1}~Hz^{-1}}$.

We find that the apparent specific luminosity exhibits a significant viewing-angle dependence in the UV and X-ray wavebands at late times, which we attribute to the structural evolution of the photosphere. Initially, the photosphere is vertically elongated, and the absence of a funnel region results in minimal mutual obscuration among different lines of sight. Consequently, the observed spectra are very similar at early times. As the accretion rate declines, the photosphere evolves into a horizontally elongated morphology. During this transition, a funnel region forms. High-energy photons generated within this funnel can be obscured when the line of sight passes through the outer layers of the photosphere, thereby creating the observed viewing-angle dependence. Furthermore, as the system evolves, the hot inner part of the photosphere becomes exposed to increasingly larger viewing angles. This gradual exposure causes the apparent specific luminosities for different lines of sight to diverge over time.

After accounting for the cosmological redshift to $z=10$, the spectral peak of the radiation is shifted to the NIR band. The peak flux exceeds $10^3\,\mathrm{nJy}$ at early times (e.g., $t \sim 110.0\,\mathrm{days}$) and remains above $10^2\,\mathrm{nJy}$ throughout most of the evolution, up to $t \sim 5489.0\,\mathrm{days}$. However, for lines of sight with extremely large inclination angles, such as $\theta = 90^\circ$, the flux decreases to the level of $10^1\,\mathrm{nJy}$ after approximately $3410.0\,\mathrm{days}$ and falls below $10\,\mathrm{nJy}$ at the end of the simulation at $t = 5489.0\,\mathrm{days}$. These flux levels indicate that Pop~III star TDEs have significant potential for detection by both JWST and the Roman Space Telescope.

The region where the observed flux exhibits a strong viewing-angle dependence is shifted into the optical/UV bands in the observer's frame. Furthermore, the late-time increase in flux observed in the high-energy bands (optical to EUV) is a direct result of the increasing exposure of the inner hot photospheric region as the system evolves. This geometric evolution alters the shape of the light curve, producing a characteristic rebrightening phase at these wavelengths.

However, we find that the flux in the high-energy wavebands (optical/UV) is suppressed to low levels under the specific extinction model adopted in this study. Consequently, the viewing-angle dependence, which is present in these bands, becomes nearly undetectable. Only a marginal signature remains observable in the NIR region during the late-time evolution between $t = 1210.0\,\mathrm{days}$ and $t = 3960.0\,\mathrm{days}$.

Finally, we calculate the radio flare produced by the wind-CNM interaction. We find that the radio luminosity increases continuously over the calculated evolution of 1490.5 days. This persistent brightening is due to the exceptionally large mass of the strong unbound wind from the super-Eddington accretion flow in Pop III star TDEs, which prevents the wind from being effectively decelerated by the CNM. After accounting for cosmological redshift, the resulting radio flux is predicted to rise over an unusually long timescale of $10^4\,\mathrm{days}$. The flux density at 1.0 GHz can exceed $10^2\,\mathrm{nJy}$ and even reach $10^3\,\mathrm{nJy}$ at late times. This suggests that such a long-lived, rising radio transient could be a promising candidate for future detection.

Our simulation results show broad agreement with the analytical study of Pop III star TDEs by \cite{ChowdhuryChang&Daietal.2024}. Both works find that the spectral energy distribution peaks in the NIR band, with flux densities exceeding $10^2~\mathrm{nJy}$, indicating a strong potential for detection with the JWST and the Nancy Grace Roman Space Telescope. However, a key distinction lies in the treatment of the system's geometry. The one-dimensional model adopted in \cite{ChowdhuryChang&Daietal.2024} lacks the necessary physics to describe the funnel region, and thus cannot produce the high energy X-ray emission or the viewing-angle dependence found in our work. Based on our radiative hydrodynamic simulations, we demonstrate that the structural evolution of the photosphere naturally produces a clear viewing-angle dependence in the high-energy bands (UV/X-ray). This result provides an advancement beyond previous analytical models by linking the observed spectral features directly to the evolving accretion flow geometry.

On the other hand, although the overall spectral shape in the infrared band is similar, the temporal evolution of the spectrum reported in \cite{ChowdhuryChang&Daietal.2024} differs from our findings. Their analytical model predicts that the spectral luminosity initially increases during the expansion phase of the photosphere. A subsequent decrease in the NIR flux occurs only after a critical time $t_{\mathrm{edge}}$, when the photosphere reaches its maximum extent and begins to recede due to the declining accretion rate. In contrast, our results show that the light curve begins to decrease before the photosphere has reached its maximum size.

This discrepancy arises from the simplifying assumption in the analytical model proposed by \cite{2009MNRAS.400.2070S} and adopted by \cite{ChowdhuryChang&Daietal.2024}, which posits a density profile at the photosphere of $\rho(R_{\mathrm{ph}}) \sim (\kappa_{\mathrm{s}} R_{\mathrm{ph}})^{-1}$, where $\kappa_{\mathrm{s}}$ is the electron scattering opacity. When combined with the temperature scaling $T \propto \rho^{1/3}$ for adiabatic expansion, this yields a luminosity evolution of $L \propto r^{2} T^{4} \propto r^{2} \rho^{4/3} \propto r^{2/3}$, predicting an increase in luminosity with an expanding photospheric radius. However, a steady-state wind obeying mass conservation follows the density profile $\rho \propto r^{-2}$, which is consistent with our simulation results. Using this relation, the luminosity scales as $L \propto r^{2} \rho^{4/3} \propto r^{-2/3}$, indicating a decrease during the adiabatic expansion phase. This earlier decline combined with the subsequent rebrightening due to the exposure of the inner hot region in our simulation thus produces a different light curve evolution compared to the previous analytical work.

Furthermore, the analytical model predicts a significant late-time increase in the spectral peak frequency due to the overall shrinkage of the photosphere and the corresponding temperature rise. However, our simulation reveals a more complex geometric evolution: while the photosphere recedes significantly along the rotational axis—affecting the high-energy emission—it remains extensively extended along the mid-plane, which dominates the peak optical/UV emission. This mid-plane region does not shrink substantially; consequently, the peak frequency at late times does not increase markedly. The distinct recession behavior of the photosphere also modifies the late-time spectral peak.

In this study, we calculate the emission from the photosphere directly based on its blackbody intensity. Other physical processes, such as (inverse) Compton scattering or photon reflection, could modify the spectral shape, particularly at high energies. However, since our emission is derived from the photosphere, the optical depth of the material traversed by photons as they propagate outward is less than unity, implying that Compton scattering is not significant. We therefore expect these effects to introduce only quantitative changes that would not alter the main conclusions of this work.

In this work, we adopted a specific extinction model. The actual extinction properties in high-redshift environments may differ from our assumptions, and thus the detailed features of the received spectrum could be revised by future studies of the absorbing material in such environments. However, the emission in the far-infrared and near-infrared bands is only weakly affected by dust extinction. Therefore, the conclusion of this work—that such Pop III star TDE systems possess strong potential for detection by both JWST and the Roman Space Telescope—remains robust.

The radio emission presented here is obtained by summing the contributions from all angular bins. This approach could, in principle, introduce an approximation, as the exact spectrum likely depends on the line of sight due to variations in synchrotron self-absorption (SSA) with angle, and thus should not be simply summed. However, given the cosmological redshift of $z=10$, the observed 1.0 GHz emission corresponds to a rest-frame frequency of 11.0 GHz. At this higher rest-frame frequency, the emission is expected to be optically thin, above the SSA frequency. Therefore, for the specific observed wavelength of interest, our summed result remains robust. Furthermore, we note that our radio calculation is based solely on the properties during the most powerful phase of the wind. Since the wind persists for over $500.0\,\mathrm{days}$, a more complete simulation that follows the long-term wind-CNM interaction represents an important direction for future work.

\section{Acknowledgments}

  D-F.B is supported by the National SKA Program of China (No.2025SKA0130100) and the Natural Science Foundation of China (grants 12192220, 12192223). L.C is supported by NSFC (12173066), National Key R$\&$D program of China (2024YFA1611403), National SKA Program of China (2022SKA0120102) and Shanghai Pilot Programme for Basic Research, CAS Shanghai Branch (JCYJ-SHFY-2021-013). B-Y.C acknowledges the support from China's Space Origins Exploration Program. X-HY is supported by Chongqing Natural Science Foundation (grant CSTB2023NSCQ-MSX0093) and the Natural Science Foundation of China (grant 12347101). Computational work was performed using high-performance computing resources at the Advanced Research Computing Core Facility of Shanghai Astronomical Observatory.

\bibliography{sample701}{}
\bibliographystyle{aasjournalv7}


\appendix

\section{RESULTS OF MODEL M300-5}

This appendix presents the emission properties derived from our simulations of model M300-5, in which we adopt a metallicity of \( Z = 10^{-5} Z_{\odot} \).  Compared to model M300-9, this model has a lower peak fallback rate and thus a lower initial accretion rate. However, the fallback timescale is longer, resulting in slower overall evolution. All figures in this section follow the same format and conventions as described in the main text.

\begin{figure}
    \centering
    \includegraphics[width=0.7\linewidth]{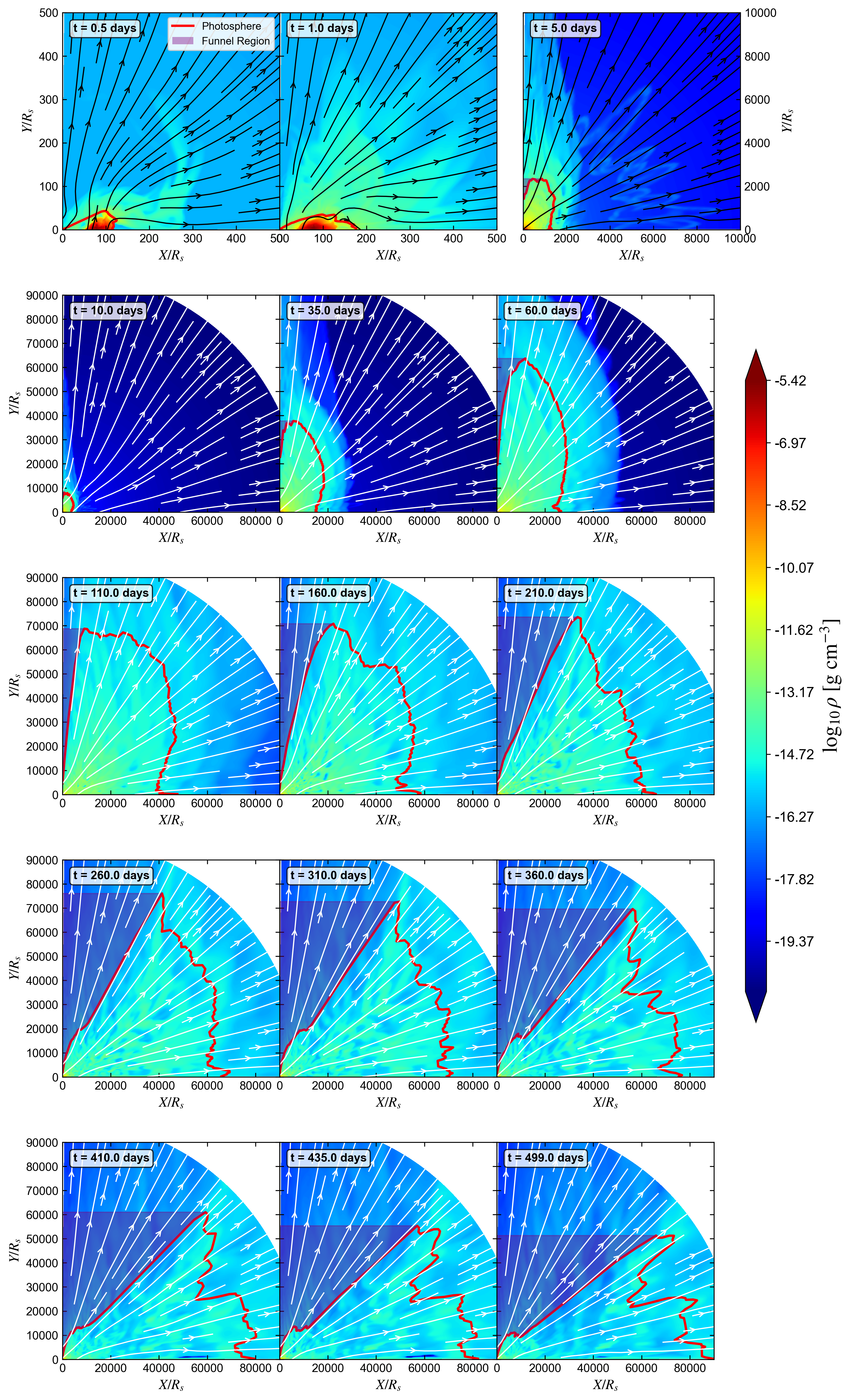}
    \caption{Snapshots of gas density (color scale) and fluid velocity streamlines (black arrows are used for the panels in the first row, and white arrows for the other panels.) for model M300-5 at 15 evolutionary stages. The data presentation follows the same convention as in Fig.~\ref{fig1}.}
    \label{figA1}
\end{figure}

\begin{figure}
    \centering
    \includegraphics[width=0.8\linewidth]{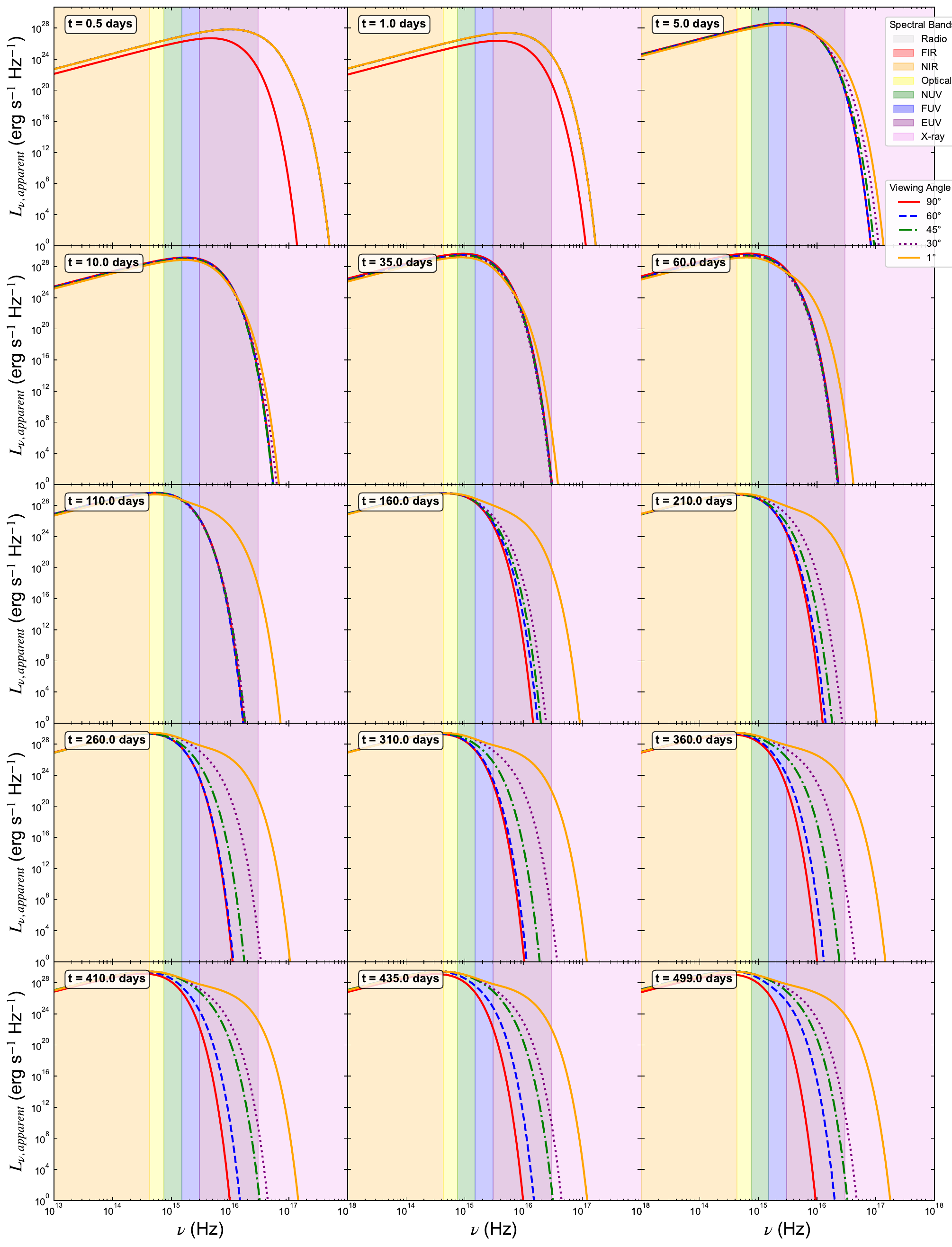}
    \caption{Temporal evolution of the apparent specific luminosity for model M300-5, shown for 15 discrete snapshots from $t = 0.5$ to $499.0\,\mathrm{days}$.The data presentation follows the same convention as in Fig.~\ref{fig4}.}
    \label{figA2}
\end{figure}
\begin{figure}
    \centering
    \includegraphics[width=0.8\linewidth]{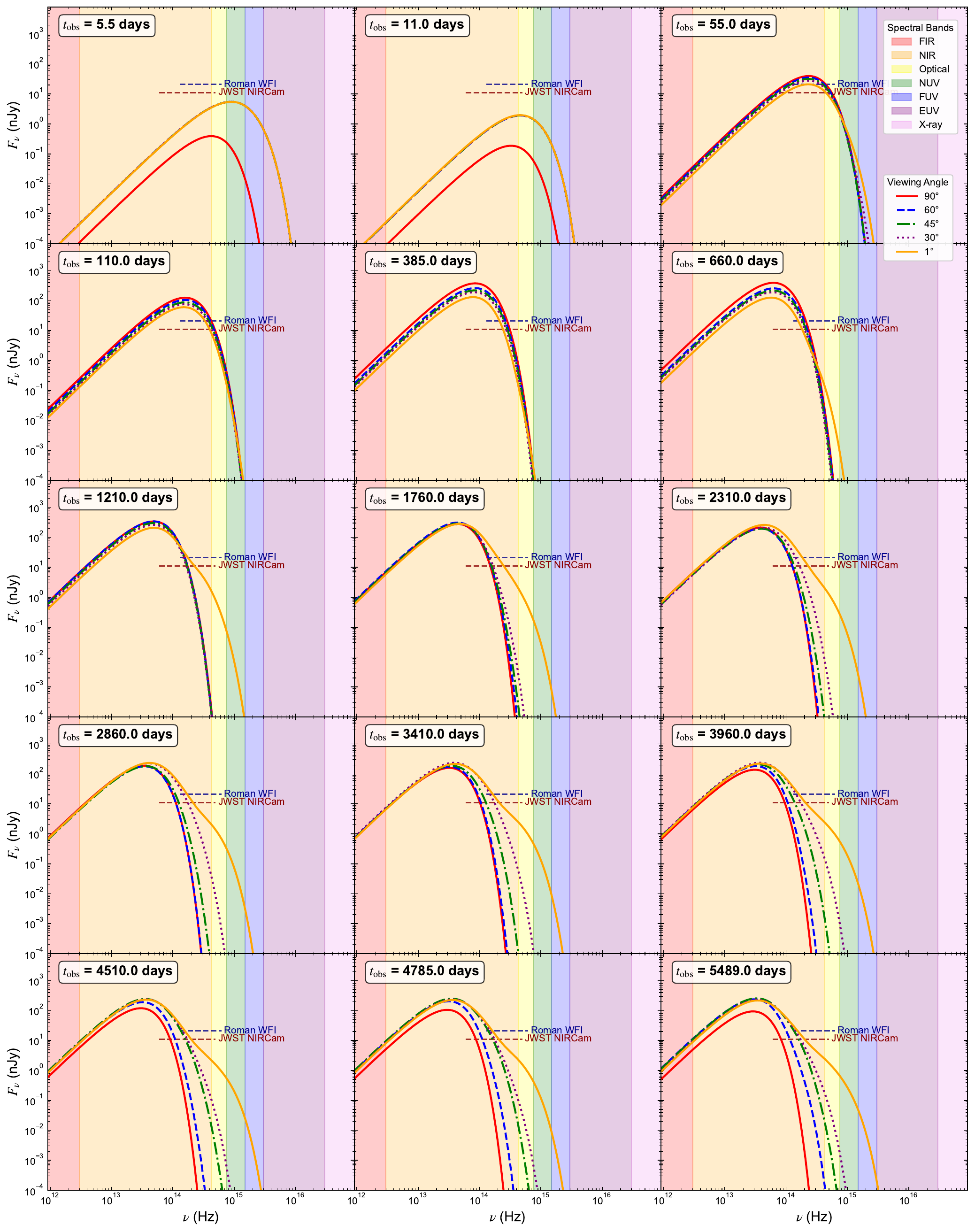}
    \caption{Temporal evolution of the flux density for model M300-5, displayed at 15 discrete snapshots from $t_{\rm obs} = 5.5$ to $5489.0\,\mathrm{days}$, where $t_{\rm obs}$ is the observer frame time. The calculation assumes a cosmological redshift of $z = 10$ and does not include extinction effects. The horizontal red and blue dashed lines indicate the detection limits of JWST and the Roman Space Telescope, respectively. The color bands (denoting spectral wavebands) and line styles (representing different viewing angles) follow the same convention as in Fig.~\ref{fig4}.}
    \label{figA3}
\end{figure}
\begin{figure}
    \centering
    \includegraphics[width=0.9\linewidth]{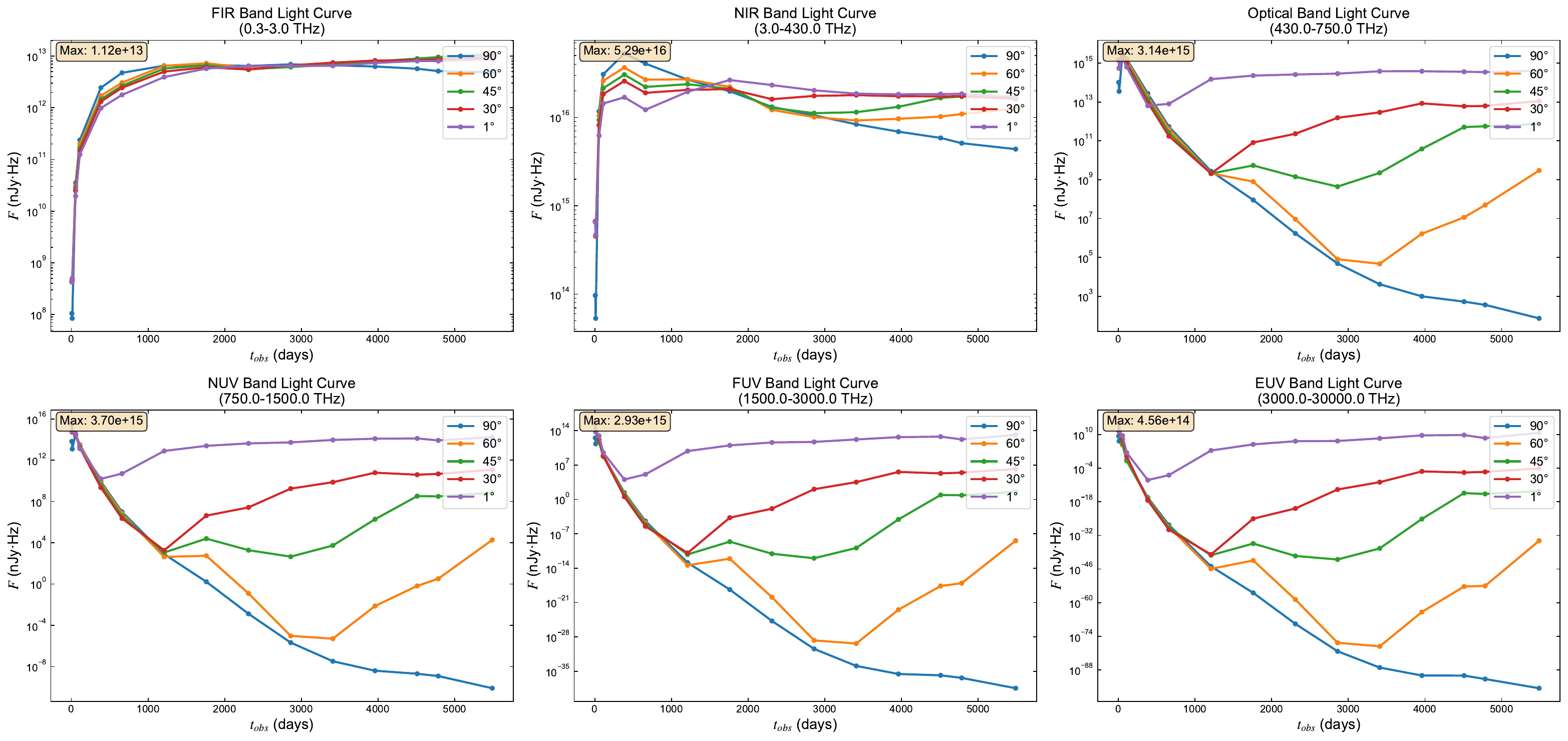}
    \caption{Received light curves in different wavebands for model M300-5, assuming a cosmological redshift of $z = 10$ and no extinction. The data presentation follows the same convention as in Fig.~\ref{fig6}.}
    \label{figA4}
\end{figure}
\begin{figure}
    \centering
    \includegraphics[width=0.8\linewidth]{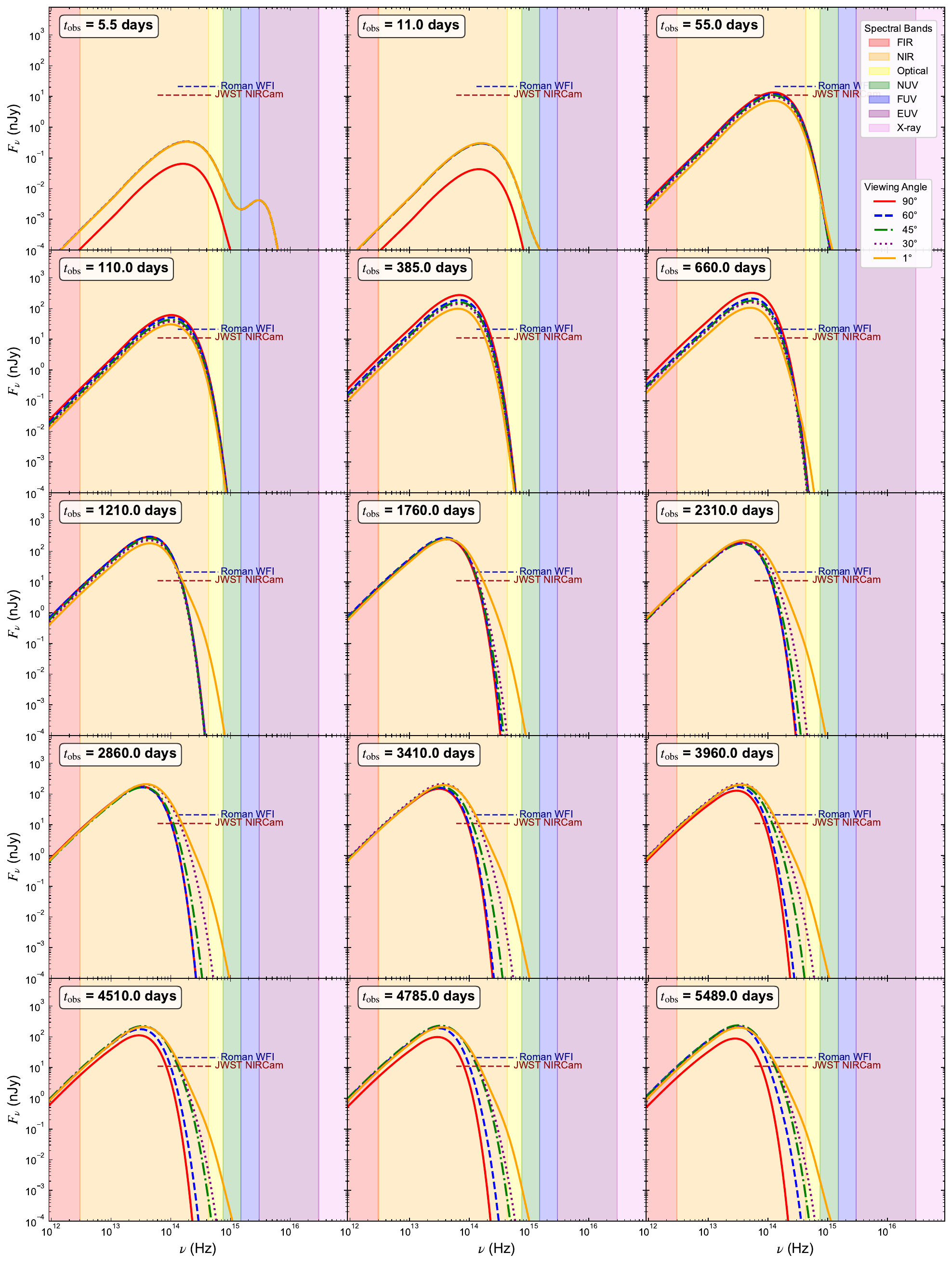}
    \caption{Received flux density for model M300-9, calculated using the extinction model described in this work. The data presentation and viewing angles follow the same convention as in Fig.~\ref{fig5}.}
    \label{figA5}
\end{figure}
\begin{figure}
    \centering
    \includegraphics[width=0.9\linewidth]{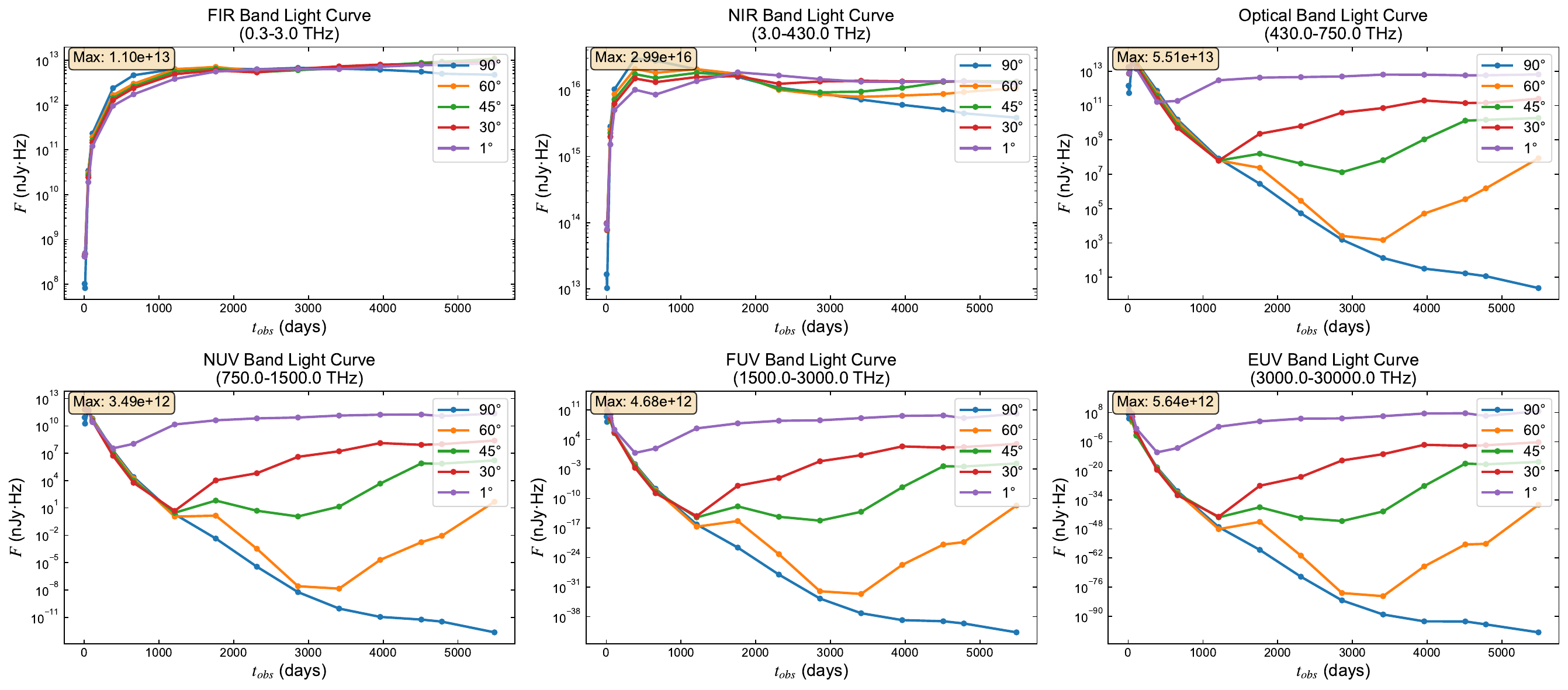}
    \caption{Received light curves in different wavebands for model M300-5, calculated using the extinction model described in this work. The data presentation and viewing angles follow the same convention as in Fig.~\ref{fig6}.}
    \label{figA6}
\end{figure}
\begin{figure}
    \centering
    \includegraphics[width=0.9\linewidth]{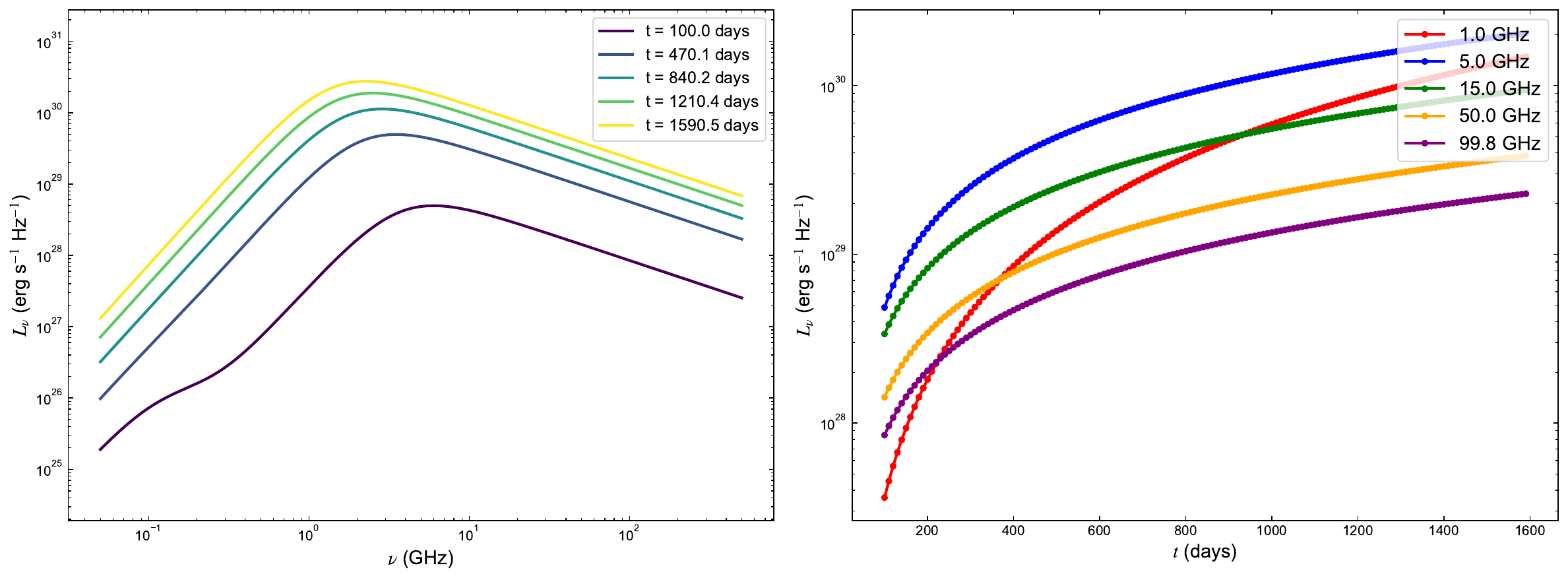}
    \caption{Radio spectral evolution for model M300-5. The left panel shows the spectra at five different times. The data presentation follows the same convention as in Fig.~\ref{fig10}.}
    \label{figA7}
\end{figure}
\begin{figure}
    \centering
    \includegraphics[width=0.9\linewidth]{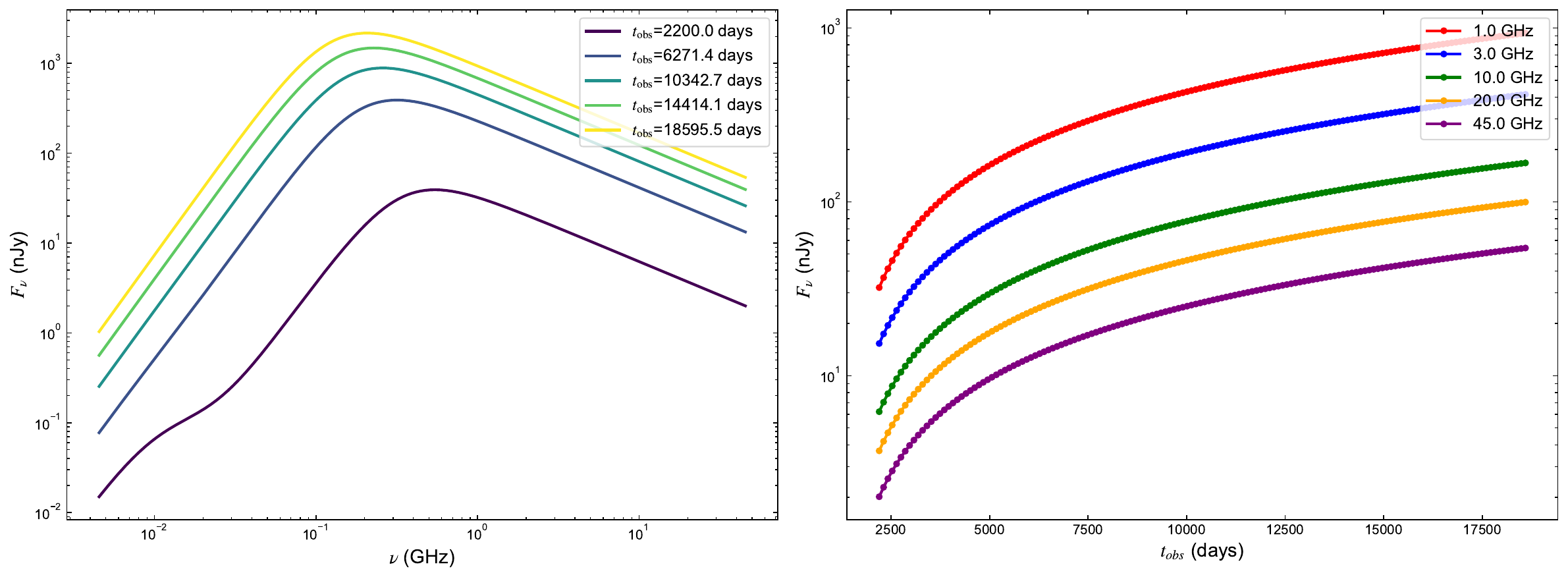}
    \caption{Observed radio flux density for model M300-9 after applying a cosmological redshift of $z=10$. The data presentation follows the same convention as in Fig.~\ref{fig11}.}
    \label{figA8}
\end{figure}

The general conclusions for this model are similar to those for M300-9, but the evolution is slower. The primary differences are as follows. First, the initial expansion of the photosphere is slower, and it maintains a horizontally elongated, 'disk-like' morphology during the early phase. Second, due to this geometry, mutual obscuration is significant from the beginning. Consequently, the specific luminosity along the $\theta = 90^\circ$ line of sight is lower than that from other viewing angles in the initial state.

For the same reason explained above, significant X-ray and EUV emission emerges at late times in Model M300-5, differing from the simplified ‘no-leak’ conclusion in our prior work (\cite{2025arXiv251221500S}).

Furthermore, owing to the longer evolutionary timescale of model M300-5, the viewing-angle dependence of the received spectrum remains significant for an extended duration even after accounting for extinction effects.

\begin{figure}
    \centering
    \includegraphics[width=1.0\linewidth]{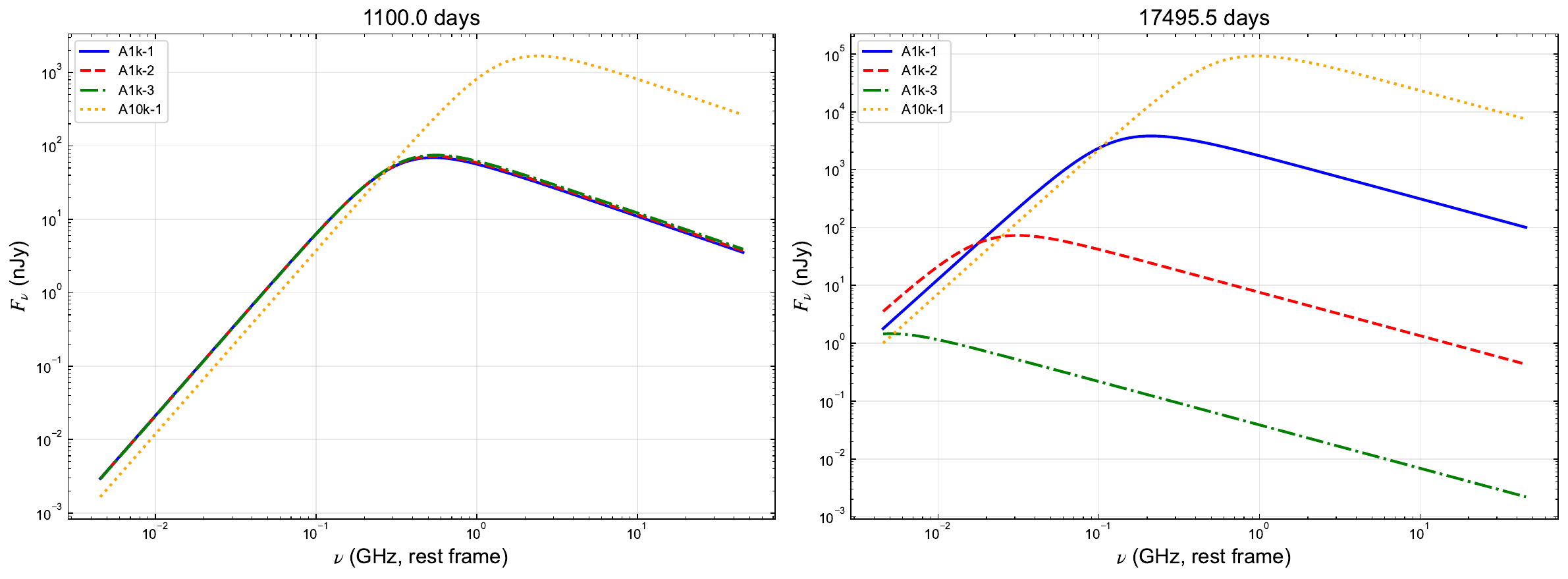}
    \caption{Specific flux of the received radio emission at $t_{\rm obs} = 110.0\,\mathrm{days}$ (left panel) and $t_{\rm obs} = 17495.5\,\mathrm{days}$ (right panel). The blue solid line shows the result for the fiducial model A1k-1. The red dashed, green dash-dotted, and yellow dotted lines represent the results for models A1k-2, A1k-3, and A10k-1, respectively.}
    \label{figB1}
\end{figure}
\begin{figure}
    \centering
    \includegraphics[width=1.0\linewidth]{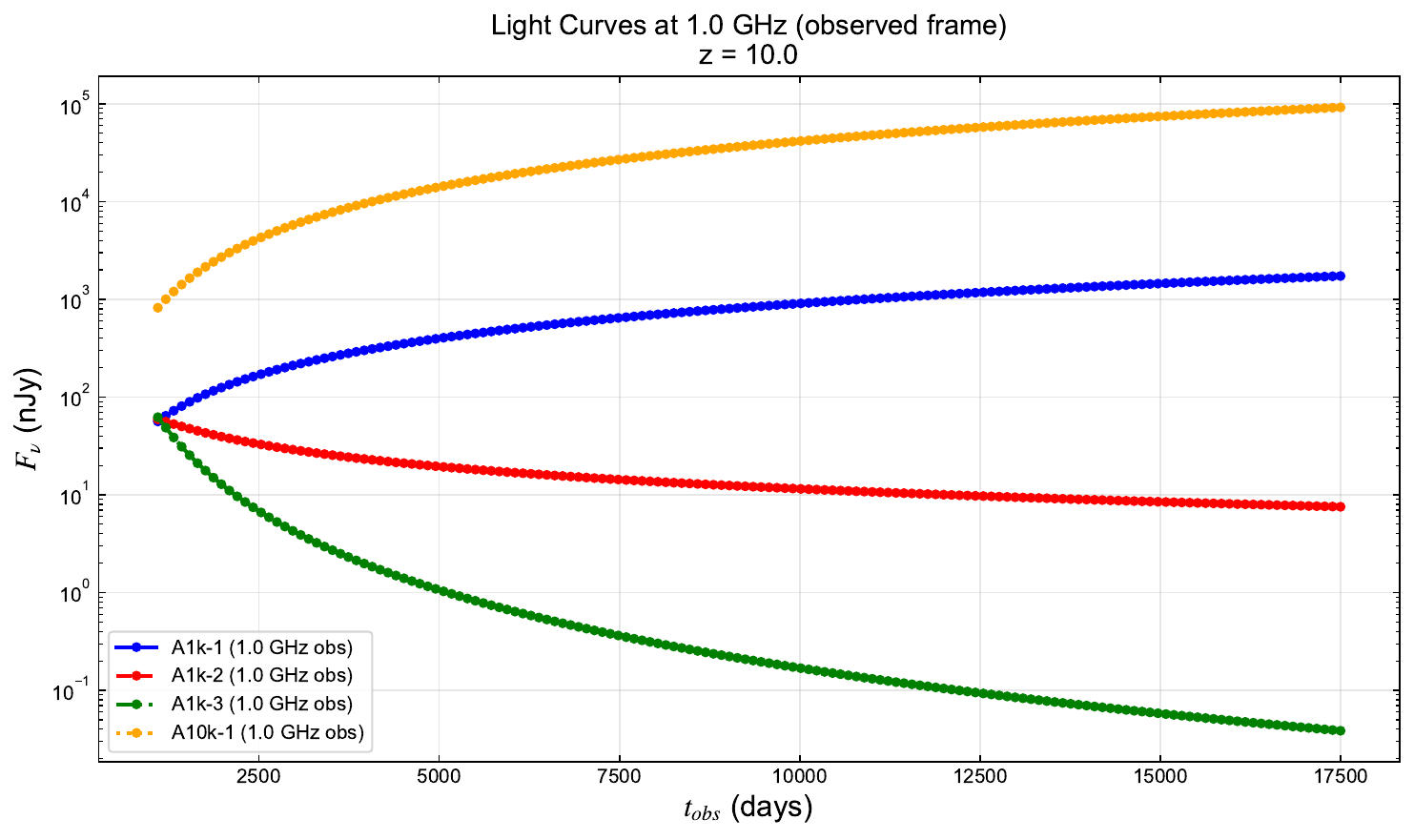}
    \caption{The light curves at an observed frequency of $\nu_{\rm obs} = 1.0\,\mathrm{GHz}$ for the different models. The blue solid line shows the result for the fiducial model A1k-1. The red, green, and yellow lines represent the results for models A1k-2, A1k-3, and A10k-1, respectively.}
    \label{figB2}
\end{figure}
\section{THE EFFECT OF CNM MODEL}
\label{Appendix.B}
This section presents the results calculated with different CNM density profiles to investigate the influence of the CNM environment on the observed emission. In this study, we assume a power-law density profile for the CNM, \( n_{\mathrm{CNM}} = A r_{-2}^{k} \, \mathrm{cm^{-3}} \), with parameters \( A = 8.5 \times 10^3 \) and \( k = -1 \) based on the fit to the TDE AT2019azh presented in \cite{2025ApJ...979..109Z}. However, the properties of the CNM, particularly its density and radial distribution, could be significantly different in high-redshift environments. The impact of adopting an alternative CNM model can be estimated analytically under the premise that the wind possesses sufficient mass to propagate with approximately constant velocity. Within this framework, a change in the CNM density profile primarily scales the key spectral properties as derived in Section~\ref{subsection3.2}: the synchrotron frequency scales as \( \nu_{\mathrm{m}} \propto n_{\mathrm{CNM}}^{1/2} \), and the specific luminosity at \( \nu_{\mathrm{m}} \) scales as \( L_{\nu_{\mathrm{m}}} \propto n_{\mathrm{CNM}}^{3/2} r^3 \). Consequently, an increase in the density normalization parameter \( A \) would lead to an increase in both the synchrotron frequency \( \nu_{\mathrm{m}} \) and the specific luminosity \( L_{\nu_{\mathrm{m}}} \). Conversely, a variation in the power-law index \( k \) alters the temporal evolution: a smaller \( k \) results in a slower temporal increase in luminosity, and if \( k < -2 \), the luminosity would instead decrease with time. 

This straightforward analytical expectation is consistent with the numerical results for the model M300-9 presented here.

Figure~\ref{figB1} shows the specific flux of the received radio emission for different CNM models, where `A1' denotes $A = 8.5 \times 10^3$, `A10' denotes $A = 8.5 \times 10^4$, and `k-x' indicates the value of the power-law index $k=-x$. Model A1k-1 is the CNM profile we adopt in the main text. Model A10k-1 exhibits a higher flux and a higher peak frequency. In contrast, Models A1k-2 and A1k-3 show a lower peak frequency at late times. Consistent with the analytical scaling $L_{\nu_{\mathrm{m}}} \propto n_{\mathrm{CNM}}^{3/2} r^3$, the flux for Model A1k-2 evolves only insignificantly with time, while that for Model A1k-3 decreases over time.

Figure~\ref{figB2} presents the light curves at an observed frequency of $\nu_{\rm obs} = 1.0\,\mathrm{GHz}$ for the different models. Model A1k-2 exhibits a slightly declining light curve, which results from the nearly constant specific luminosity at $\nu_{\rm m}$ combined with a decreasing $\nu_{\rm m}$. This behavior is also consistent with the simple analytical expectation.

\end{document}